\documentclass[10pt]{article}

\pdfoutput=1 % if your are submitting a PDFlatex (i.e. if you have
\usepackage{jheppub} % for details on the use of the package, please
\usepackage[T1]{fontenc} % if needed
\usepackage{enumitem}
\usepackage{xcolor}
\usepackage{amssymb,amsthm,cancel,hyperref,graphicx,xcolor}
\usepackage{mathrsfs,wasysym}
\usepackage{booktabs}
\usepackage{subcaption}
\usepackage{tikz}
\usetikzlibrary{calc}
\usepackage{placeins}
\usepackage{tabu}
\usepackage{pdflscape}
\usepackage{afterpage}
\newcommand{\as}{\alpha_s}

\let\originalleft\left
\let\originalright\right
\renewcommand{\left}{\mathopen{}\mathclose\bgroup\originalleft}
\renewcommand{\right}{\aftergroup\egroup\originalright}

\newcommand{\zc}{z_{\rm cut}}
\newcommand{\zcb}{z_\beta}

\newcommand{\herwig}{\textsf{Herwig}}
\newcommand{\pythia}{\textsf{Pythia}}
\newcommand{\sherpa}{\textsf{Sherpa}}
\newcommand{\dire}{\textsf{Dire}}
\newcommand{\rivet}{\textsf{Rivet}}
\newcommand{\openloops}{\textsf{OpenLoops}}
\newcommand{\fastjet}{\textsf{FastJet}}

\newcommand{\eventtwo}{\textsf{EVENT2 }}

\def\beq{\begin{equation}}  
\def\eeq{\end{equation}}
\def\({\left(}
\def\){\right)}
\def\[{\left[}
\def\]{\right]}

\preprint{\newline FERMILAB-PUB-19-266-T\\MCNET-19-14}

\title{\boldmath Fitting the Strong Coupling Constant with Soft-Drop Thrust}

\author[a]{Simone~Marzani,} 
\author[b,c]{Daniel~Reichelt,}
\author[b]{Steffen~Schumann,}
\author[d]{Gregory~Soyez,}
\author[b,d]{Vincent~Theeuwes.}

\affiliation[a]{Dipartimento di Fisica, Universit\`a di Genova and INFN, Sezione di Genova,\\ Via Dodecaneso 33, 16146, Genoa, Italy}
\affiliation[b]{Institut f{\"u}r Theoretische Physik, Georg-August-Universit{\"a}t G\"ottingen,\\ Friedrich-Hund-Platz 1, 37077 G\"ottingen, Germany}
\affiliation[c]{Fermi National Accelerator Laboratory, Batavia, IL, 60510-0500, USA}
\affiliation[d]{Institut de Physique Th\'eorique, Paris Saclay University, CEA, CNRS, F-91191 Gif-sur-Yvette France}
\emailAdd{simone.marzani@ge.infn.it}
\emailAdd{daniel.reichelt@uni-goettingen.de}
\emailAdd{steffen.schumann@phys.uni-goettingen.de}
\emailAdd{gregory.soyez@ipht.fr}
\emailAdd{vincent.theeuwes@uni-goettingen.de}

\abstract{Soft drop has been shown to reduce hadronisation effects at $e^+e^-$ colliders for the thrust event shape. In this context, we perform fits of the strong coupling constant for the soft-drop thrust distribution at NLO+NLL accuracy to pseudo data generated by the \sherpa~event generator. In particular, we focus on the impact of hadronisation corrections, which we estimate both with an analytical model and a Monte-Carlo based one, on the fitted value of $\alpha_s(m_Z)$. We find that grooming can reduce the size of the shift in the fitted value of $\as$ due to hadronisation.
In addition, we also explore the possibility of extending the fitting range down to significantly lower values of (one minus) thrust. Here, soft drop is shown to play a crucial role, allowing us to maintain good fit qualities and stable values of the fitted strong coupling.
The results of these studies show that soft-drop thrust is a promising candidate for fitting $\as$ at $e^+ e^-$ colliders with reduced impact of hadronisation effects.
}

\usepackage{graphicx}
\begin{document}
\maketitle

\section{Introduction}

The CERN Large Hadron Collider (LHC) has recently finished its second physics run and it has entered a long shutdown phase,
which is devoted to upgrades of the main experiments. Meanwhile, physics analyses employing the full dataset are being carried
out, leading to results with astonishing experimental precision. The absence of any clear indications of new particles or
interactions necessitates more and more detailed comparisons of experimental measurements with theoretical predictions.
Correspondingly, high-accuracy calculations are of crucial importance in order to match the theoretical uncertainty to the
ever decreasing experimental one and to gain sensitivity even to tiny deviations from the Standard Model. In this context,
the theory community has put huge efforts in improving perturbative calculations in QCD, both at fixed-order
and at the resummed level. This includes the development of sophisticated--and highly automated--simulation tools that turn
high-precision calculations into explicit predictions for the complex final states produced in LHC proton-proton collisions.  

A crucial quantity entering all these computations is the strong coupling constant $\as$, which needs to be known at high
precision. To give an example, even the leading-order contribution to Higgs-boson production in gluon fusion at the LHC
only starts at $\mathcal{O}(\as^2)$. The current value of the coupling constant as determined by the Particle Data Group is
$\as(m_Z)=0.1181 \pm 0.0011$~\cite{Tanabashi:2018oca}, determined as the average of various different $\as$ extractions. 
This average is dominated by precise fits from lattice QCD~\cite{Maltman:2008nf,Aoki:2009tf,McNeile:2010ji,Blossier:2012ef,Chakraborty:2014aca,Bazavov:2014wgs,Zafeiropoulos:2019flq}, followed by measurements of event-shape variables at $e^+e^-$
colliders~\cite{Bethke:2008hf,Dissertori:2009ik,Dissertori:2009qa,OPAL:2011aa,Schieck:2012mp,Davison:2008vx,Abbate:2010xh,Gehrmann:2012sc,Hoang:2014wka}. 

The most accurate $\as$ determination from event shapes has been obtained by fitting the thrust distribution~\cite{Farhi:1977sg}
to a theoretical calculation of outstanding precision, namely next-to-next-to-leading order matched to a resummed calculation
which accounts for the first three towers of logarithmic contributions, i.e.\ NNLO+N$^3$LL accuracy~\cite{Abbate:2010xh,Abbate:2012jh}. 
Noticeably, this result ($\as(m_Z)=0.1135 \pm 0.0011$) exhibits some tension with the world average. An analogous analysis
performed for the C-parameter~\cite{Hoang:2014wka} resulted in a similar deviation. A common feature of these extractions is
a sizeable non-perturbative contamination originating from the observed final states being composed of hadrons, rather than
partons. The leading contribution to this component can be modelled analytically in terms of a single non-perturbative
parameter $\Omega$, which is fitted jointly with the strong coupling. Unfortunately, $\as$ and $\Omega$ are highly
correlated. It would thus be beneficial to break this degeneracy in
order to ensure that the extraction of $\as$ is less dependent on
non-perturbative physics.
%unbiased. 

In recent studies~\cite{Baron:2018nfz,Bendavid:2018nar}, some of us put forward the idea of exploiting techniques developed in jet physics to
improve on the determination of the strong coupling. Jet substructure techniques are primarily developed to search for boosted
massive particles decaying hadronically. However, extensive literature
on the topic (see for instance Refs.~\cite{Marzani:2019hun,Larkoski:2017jix,Asquith:2018igt}
for recent reviews) has demonstrated their ability to extend the range of applicability of perturbative methods in jet physics.
This essentially originates from a reduced sensitivity to very soft emissions, resulting in milder hadronisation corrections.
The soft-drop algorithm~\cite{Larkoski:2014wba} is particularly well-suited for this task and in Ref.~\cite{Baron:2018nfz}
soft-dropped versions of thrust, and related event shapes, where discussed in detail. In this paper we continue the analysis
of soft-drop thrust and, in particular, focus on the effects that soft drop has
on the impact of non-perturbative corrections in the extraction of
$\as$. Furthermore, 
we investigate how the fit value and quality change as a function of the observable range considered for the extraction.
In the absence of experimental measurements of soft-drop thrust, we perform our analysis on pseudo-data simulated with
the \sherpa\ event generator~\cite{Bothmann:2019yzt,Gleisberg:2008ta}. We will illustrate the reduced sensitivity of the soft-drop thrust
observable to non-perturbative contributions. To this end we consider hadronisation corrections from Monte Carlo simulations
as well as an analytical model. This qualifies soft-drop thrust, and
jet substructure observables in general, as attractive candidates for future extractions
of $\as$. 

Our paper is organised as follows, in Sec.~\ref{sec:SD-thrust} we present the soft-drop thrust observable and some detail
on its evaluation to NLO+NLL accuracy. In Sec.~\ref{sec:MC} we discuss the generation of the pseudo data for the $\as$
fits. Our methods to account for hadronisation corrections are presented in Sec.~\ref{sec:had_models}. The fit procedure
and our treatment of statistical and systematic uncertainties get illustrated in Sec.~\ref{sec:fit-as}. The corresponding
results obtained are discussed in Sec.~\ref{sec:results}. We draw conclusions and give a brief outlook in Sec.~\ref{sec:conclusion}.

\section{Soft-drop thrust}\label{sec:SD-thrust}

Soft drop~\cite{Larkoski:2014wba} is an example of so-called grooming techniques. These are designed in the context
of jet substructure with the aim to clean up jets from soft wide-angle radiation.
The soft-drop algorithm recursively de-clusters the angular-ordered tree of a Cambridge-Aachen jet~\cite{Dokshitzer:1997in,Wobisch:1998wt}
of original radius $R$ until a hard splitting is found. The $e^+e^-$ version of soft drop uses hemisphere jets clustered
using the $e^+e^-$ version of the Cambridge-Aachen algorithm.

\noindent
For each (hemisphere) jet the method proceeds as follows:
\begin{enumerate}
	\item undo the last step in clustering thereby splitting the jet into two subjets $i$ and $j$;
	\item check if the subjets pass the soft-drop condition:
	\begin{equation}
	\frac{\min\[E_i,E_j\]}{E_i+E_j}>\zc \(1-\cos\theta_{ij}\)^{\beta/2},
	\end{equation}
	where $E_{i/j}$ denote the energies of the subjets, $\theta_{ij}$ is the angle between them and $\zc$ and $\beta$
        are parameters of the soft-drop algorithm;
      \item if the splitting fails the soft-drop condition, the softer
        subjet is discarded (groomed away) and the steps are repeated
        for the resulting jet (the harder subjet);
      \item if instead the subjets pass the condition, the procedure is terminated and the resulting jet is the
        combination of subjets $i$ and $j$.
\end{enumerate}
The soft-drop algorithm features two parameters: $\zc$ and $\beta$. The first determines how stringent the cut on the
subjet energies is, whereas the latter provides an angular suppression to grooming. While $\beta\to \infty$ corresponds
to no grooming, for $\beta=0$ no angular dependence is taken into account and the soft-drop algorithm reduces
to the modified Mass-Drop Tagger (mMDT)~\cite{Butterworth:2008iy,Dasgupta:2013ihk}.
For practical purpose, we have implemented the above procedure using
\fastjet~\cite{Cacciari:2011ma} for the jet clustering and additional manipulations.

The event shape thrust~\cite{Farhi:1977sg} is defined as
\begin{equation}
  T\equiv\max_{\vec{n}}\left(\frac{\sum_{i\in\mathcal{E}}\left|\vec{n}\cdot\vec{p}_{i}\right|}{\sum_{i\in\mathcal{E}}\left|\vec{p}_{i}\right|}\right),
\end{equation}
where $\vec{p}_{i}$ labels the three-momentum of particle $i$ and the sum extends over all particles in the
event $\mathcal{E}$. The resulting vector $\vec{n}$ defines the thrust axis. Often the related variable
\begin{equation}
\tau\equiv 1-T=\min_{\vec{n}}\left(1-\frac{\sum_{i\in\mathcal{E}}\left|\vec{n}\cdot\vec{p}_{i}\right|}{\sum_{i\in\mathcal{E}}\left|\vec{p}_{i}\right|}\right)
\end{equation}
is used instead. Henceforth, in the following we will refer to $\tau$ as thrust. 

The soft-drop thrust shape is defined following the procedure given in~\cite{Baron:2018nfz}:
\begin{enumerate}
        \item determine the thrust axis for the full final state, i.e. without any grooming;
        \item split the event into two hemispheres based on the thrust axis;
	\item apply the soft-drop procedure on each hemisphere;
	\item compute the thrust value for each groomed hemisphere
          separately, using the groomed hemisphere-jet momenta as
          reference axes.
\end{enumerate}
The resulting value for soft-drop thrust is given by:
\begin{equation}\label{eq:sd-thrust}
\tau_{\text{SD}}=\frac{\sum_{i\in\mathcal{E}_{\text{SD}}}\left|\vec{p}_i\right|}{\sum_{i\in\mathcal{E}}\left|\vec{p}_i\right|}\left[1-\frac{\sum_{i\in\mathcal{H}^{L}_{\text{SD}}}\left|\vec{n}_L\cdot\vec{p}_i\right|+\sum_{i\in\mathcal{H}^{R}_{\text{SD}}}\left|\vec{n}_R\cdot\vec{p}_i\right|}{\sum_{i\in\mathcal{E}_{\text{SD}}}\left|\vec{p}_i\right|}\right],
\end{equation}
where $\vec{n}_{L/R}$ denotes the axis for the left and right hemisphere, respectively, and the sums extend over all
particles in the full event ($\mathcal{E}$), the soft-dropped event ($\mathcal{E}_{\text{SD}}$) or the hemispheres
($\mathcal{H}_{L/R}$). 

An important difference between this definition and what was presented in the first version of
Ref.~\cite{Baron:2018nfz} is the rescaling factor in front of the observable. This additional factor
ensures collinear safety for the case $\beta=0$. The issue occurs for an event with multiple particles
in one hemisphere and a single particle in the other one. While a virtual correction will not alter the
observable value, a collinear real emission off the lone particle, that is soft enough to be groomed away,
might alter the value of $\tau_{\text{SD}}$ if the factor is not taken into account. If there are multiple
particles present in the second hemisphere the collinear emission is protected from grooming by the
clustering history. Furthermore, if $\beta>0$ the collinear emission will not be groomed away due to the
angular suppression given in the soft-drop condition. Further details on this issue are given in
Appendix~\ref{sec:col-safe}.

\subsection*{NLO+NLL resummed predictions}
In Ref.~\cite{Baron:2018nfz} the resummation of soft-drop thrust at NLL accuracy has been presented.
The calculation is based on the factorisation of the differential distribution in hard, soft and
collinear pieces, derived using Soft Collinear Effective Theory~\cite{Frye:2016aiz}. In the limit
$\tau \ll \zc \ll 1$ the differential cross section can be written as
\begin{equation}
  \frac{d\sigma}{d\tau_{\text{SD}}} = H(Q)S_G(\zc,\beta)\left[S_C(\tau_{\text{SD}},\zc,\beta)\otimes J(\tau_{\text{SD}})\right]^2\,.
\end{equation}
Here $H$ denotes the hard function, depending on the energy scale $Q$ only, $S_G$ is a global soft function
accounting for soft wide-angle emissions, $S_C$ describes soft-collinear emissions, and $J$ denotes
the jet function encapsulating the effect of hard-collinear radiation. Some detail on the various components
and in particular their one-loop, i.e. NLL expressions are collected in App.~\ref{sec:resummation}.
In \cite{Baron:2018nfz} the resummed predictions were matched to the full NLO QCD result. NNLO QCD results
for the original version of the observable definition have been presented in~\cite{Kardos:2018kth}. For our
study here we have adjusted the calculation from \cite{Baron:2018nfz} to the collinear-safe
observable definition, resulting in an NLO+NLL accuracy of our predictions.

For the resummation contribution the approximation $\zc\ll1$ is used and no finite $\zc$ corrections are
included. Finite $\zc$ effects are power corrections in $\tau$ for $\beta>0$, however for $\beta=0$ these are only power
corrections in $\zc$ (contributing at the leading-logarithmic accuracy
in $\tau$). Despite this fact we will still study values of $\zc$ as high as $0.33$. Here the
finite $\zc$ effects will only be taken into account through means of
matching to fixed order.\footnote{Although the study was limited to
  $\zc=0.1$, Ref.~\cite{Marzani:2017mva} showed that after matching to
  NLO the residual finite $\zc$ corrections were very small.} Fixed-order
corrections are computed with the publicly available program \texttt{EVENT2}~\cite{Catani:1996jh,Catani:1996vz}. 
To further validate the resummed calculation, we evaluated the soft-drop thrust observable in the
CAESAR formalism~\cite{Banfi:2004yd}, using an independent implementation in the \sherpa\
framework~\cite{Gerwick:2014gya}. In addition a separate calculation
was performed, with slightly different treatment of the resummation
uncertainty. This is further detailed in Appendix~\ref{sec:unc}.

To match the resummed result to the exact NLO QCD matrix element, i.e. the three-parton process at NLO accuracy,
we consider two different matching schemes: multiplicative and LogR matching~\cite{Banfi:2010xy}. In both
cases we consider the cumulative distribution
\begin{equation}
\Sigma(\tau_{\text{SD}}) = \frac{1}{\sigma_0}\int\limits_{0}^{\tau_{\text{SD}}}d\tau'_{\text{SD}}\frac{d\sigma}{d\tau'_{\text{SD}}}\,.
\end{equation} 
Multiplicative matching is defined by
\begin{eqnarray}
  \Sigma\(\tau_{\text{SD}}\)&=&\(\frac{C\;\Sigma_{\rm FO}}{\Sigma_{\rm exp}}\)_{\rm FO}\frac{\Sigma_{\rm res}}{C}\,,
\end{eqnarray}
where $\Sigma_{\rm FO}$ denotes the fixed-order cumulant distribution, $\Sigma_{\rm exp}$ the power-series expansion of the resummed
cumulant $\Sigma_{\rm res}$, and $C$ corresponds to $\Sigma_{\rm res}$ and $\Sigma_{\rm exp}$ evaluated at the kinematic
end-point $\tau_{\rm max}$. Aiming for NLO+NLL accuracy, $C=1$ holds and we need to expand the fixed-order ratio to ${\cal{O}}(\as^2)$ 
\begin{eqnarray}
\Sigma\(\tau_{\text{SD}}\)&=&\[1+\frac{\as}{\pi}\(\Sigma_{\rm FO}^{(1)}-\Sigma_{\rm exp}^{(1)}\)+\frac{\as^2}{\pi^2}\(\Sigma_{\rm FO}^{(2)}-\Sigma_{\rm exp}^{(2)}+\Sigma_{\rm exp}^{(1)}\left\{\Sigma_{\rm exp}^{(1)}-\Sigma_{\rm FO}^{(1)}\right\}\)\]\Sigma_{\rm res}\,.
\end{eqnarray}

\noindent
As an alternative scheme LogR matching is used
\begin{eqnarray}
\log\Sigma\(\tau_{\text{SD}}\)&=&\(\log\[\frac{C\;\Sigma_{\rm FO}}{\Sigma_{\rm exp}}\]\)_{\rm FO}+\log\[\frac{\Sigma_{\rm res}}{C}\]\\
&=&\[\frac{\as}{\pi}\(\Sigma_{\rm FO}^{(1)}-\Sigma_{\rm exp}^{(1)}\)+\frac{\as^2}{\pi^2}\(\Sigma_{\rm FO}^{(2)}-\Sigma_{\rm exp}^{(2)}+\frac{1}{2}\left\{\(\Sigma_{\rm exp}^{(1)}\)^2-\(\Sigma_{\rm FO}^{(1)}\)^2\right\}\)\]+\log\Sigma_{\rm res}, \nonumber
\end{eqnarray}
resulting in 
\begin{equation}
\Sigma\(\tau_{\text{SD}}\)=\Sigma_{\rm res}\exp\[\frac{\as}{\pi}\(\Sigma_{\rm FO}^{(1)}-\Sigma_{\rm exp}^{(1)}\)+\frac{\as^2}{\pi^2}\(\Sigma_{\rm FO}^{(2)}-\Sigma_{\rm exp}^{(2)}+\frac{1}{2}\left\{\(\Sigma_{\rm exp}^{(1)}\)^2-\(\Sigma_{\rm FO}^{(1)}\)^2\right\}\)\].
\end{equation}
Multiplicative matching will be the default choice and the variation between the two will be included in the theoretical uncertainty,
cf. Sec.~\ref{sec:fit-as}. In order to ensure that the differential cross section for the resummation and expansion vanishes
for the fixed-order kinematical end-point, the resummation is modified in accordance with~\cite{Catani:1992ua,Jones:2003yv}, cf.
App.~\ref{sec:endpoint} for details. 

\subsection*{Transition-point treatment}

We briefly want to discuss the treatment of the transition point that marks the boundary between the soft-drop regime
of the thrust distribution and the ungroomed one. Let us consider a LO configuration, where a gluon is emitted off the quark-antiquark
pair. For sufficiently large values of thrust, i.e.\ above the transition point $\tau_{\text{SD}} =\zc/2+\mathcal{O}\(\zc^2\)$, the emission 
is always hard enough not to be impacted by grooming. Above this transition point the LO distribution for soft-drop thrust coincides with
plain thrust. We note that for $\zc=1/3$ and $\beta=2$, at leading
order, the transition point coincides with the kinematic end-point,
meaning the soft-drop thrust calculation stretches over the full LO distribution. The all-order calculation also features a
transition point at $\zc/2+\mathcal{O}\(\zc^2\)$, while if we consider higher orders in the fixed-order expansion, this transition
point smears out due to the multiple emissions.

In the study of Ref.~\cite{Baron:2018nfz}, the transition point for the all-order calculation was taken as described above. However,
this had the undesirable consequence of introducing transition-point corrections that break factorisation between the global soft and soft-collinear functions. In this current analysis we
exploit the fact that at NLL accuracy we can treat the transition point between soft-drop and plain thrust as being located at
$\zc 2^{\beta/2}$, which corresponds to the transition point for soft collinear emissions. The transition point is only located at
$\zc/2$ for wide-angle emissions. Since the cross term between logarithms of 2 and logarithms of soft wide-angle origin starts
contributing at NNLL accuracy, we can make use of this treatment at NLL accuracy, thus avoiding the aforementioned complications.
In addition to this change in transition point, we can introduce a transition-point uncertainty. If the resummation uncertainty ($x_L$) is only included in logarithms of thrust and not in the logarithms of $\zc$, the transition point gains a resummation uncertainty dependence. The details on this approach are presented in Appendix~\ref{sec:unc}.

We also consider a further effect which was not taken into account in
the analysis of Ref.~\cite{Baron:2018nfz}, namely the non-trivial
interplay between multiple-emission corrections and the transition
point. Let us consider the emission of two gluons in the transition
region (kept by the grooming procedure) such that the resulting $\tau$
is above the transition point. Corrections can appear when (one of)
the individual contributions of these two emissions to the total
$\tau$ value is below the transition point. This formally NNLL
correction becomes parametrically relevant close to the transition
point and is calculated in detail in
Appendix~\ref{sec:mult-trans}.\footnote{This correction cures the
  discontinuity in NLL distributions at the transition
  point noted out e.g.\ in~\cite{Larkoski:2014wba,Baron:2018nfz}.}

\begin{figure}[h!]
	\centering
	\includegraphics[width=\textwidth]{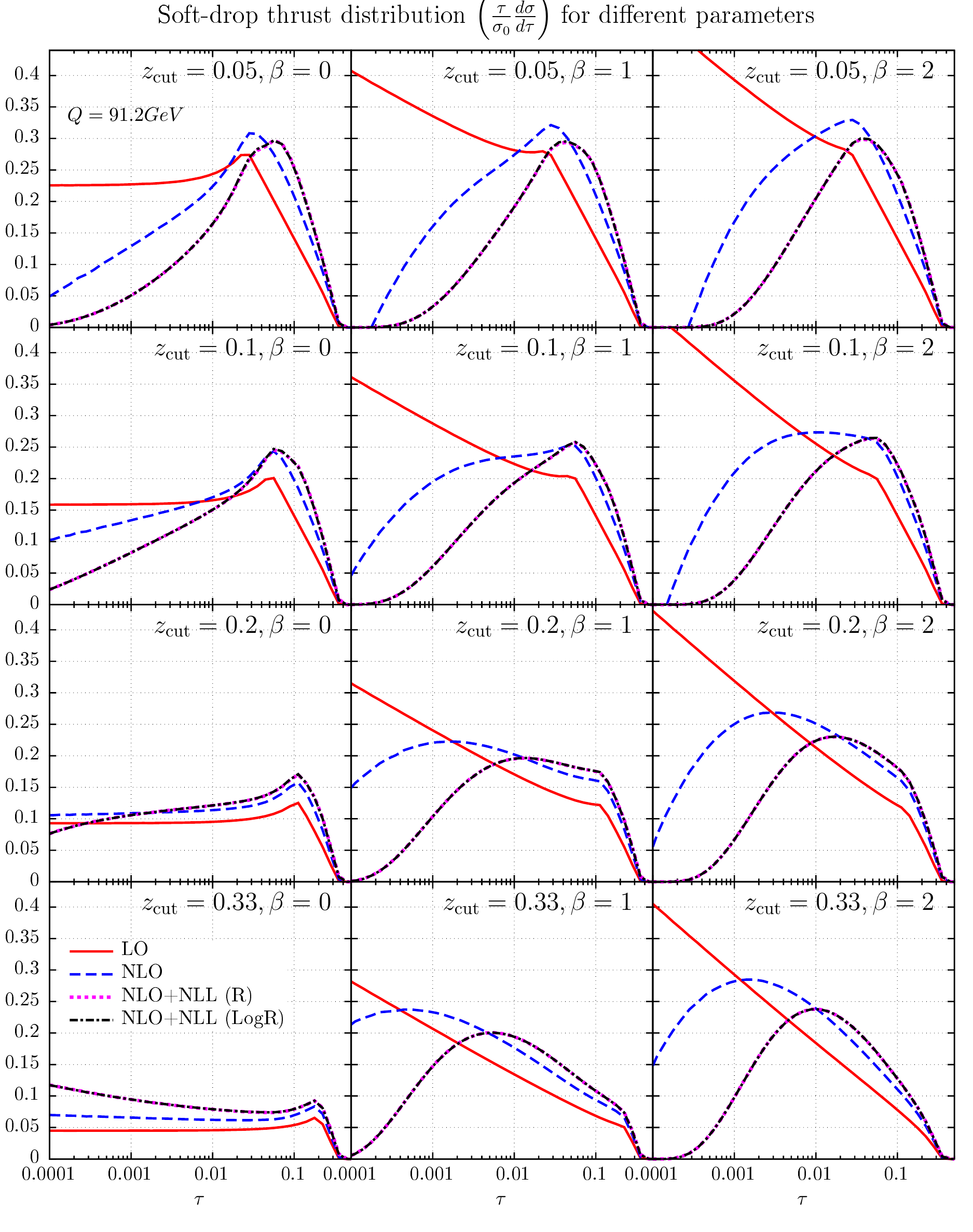}
	\caption{The differential cross section for the analytical
		soft-drop thrust distribution at LO (solid red), NLO (dashed
		blue) and NLO+NLL with both multiplicative  (dotted magenta)
		and LogR (black dashed-dotted) matching.}\label{fig:Analytical_SD}
\end{figure}

In Figure~\ref{fig:Analytical_SD} the results of these analytical computations are presented. From left to right the value of $\beta$ is varied between \{0, 1, 2\}, whereas top to bottom shows different values of $\zc=$\{0.05, 0.1, 0.2, 0.33\}. The results are presented at LO, NLO and NLO+NLL using both matching schemes. First a good agreement between the two different matching schemes can be seen. Both the contributions from the fixed-order transition point near $\zc/2$ and the resummation transition point can be seen. For lower values of $\zc$ there is still a significant contribution from the second transition point, which is significantly reduced for larger values of $\zc$.

It is interesting to point out that
soft-drop also affects the thrust distribution above the transition
point. We have already talked about transition point corrections in
the context of our resummed calculation (see also
Appendix~\ref{sec:mult-trans}). For a fixed-order calculation the
soft-drop condition impacts the thrust distribution at all values of
thrust. Finally, additional soft emissions, such as the ones related
to the non-perturbative model, are also affected by grooming even
above the transition point.

\clearpage

\section{Monte-Carlo based pseudo data}\label{sec:MC}

Given that we attempt to extract the strong coupling from an observable where there exists no
actual measurement yet, we have to resort to simulated data. Furthermore, we use Monte Carlo
simulations to assess hadronisation corrections and the associated uncertainties. 

We employ the \sherpa\ event generator~\cite{Gleisberg:2008ta,Bothmann:2019yzt}
version 2.2.5 to simulate $e^+e^-\to\textrm{hadrons}$ events at LEP1 energy of $\sqrt{s}=m_Z$. We generate
the hard-scattering configurations for varying parton-multiplicity final states at next-to-leading
order in QCD. The matrix elements get matched to the \sherpa\ dipole parton shower~\cite{Schumann:2007mg}
based on the \sherpa\ implementation of the MC@NLO method ~\cite{Frixione:2002ik,Hoeche:2011fd} and
merged into inclusive samples according to the MEPS@NLO formalism~\cite{Hoeche:2012yf}. The required one-loop
virtual amplitudes are obtained from the \openloops-1.3.1~\cite{Cascioli:2011va} package. The default
hadronisation model in \sherpa\ is the cluster fragmentation described in~\cite{Winter:2003tt}. However, \sherpa\
also provides the option to invoke the Lund string fragmentation~\cite{Sjostrand:1982fn} as
implemented in \pythia\ 6.4~\cite{Sjostrand:2006za}. Using the identical perturbative inputs, i.e.
shower-evolved events, this provides us with a consistent estimate for the hadronisation related
uncertainties. To analyse events we employ the \rivet-2.7.2 package~\cite{Buckley:2010ar}. To this end
we have implemented the soft-drop thrust observable using \fastjet-3.3.2~\cite{Cacciari:2011ma} for
particle clustering.

\begin{figure}
  \centering
  \includegraphics[width=.5\textwidth]{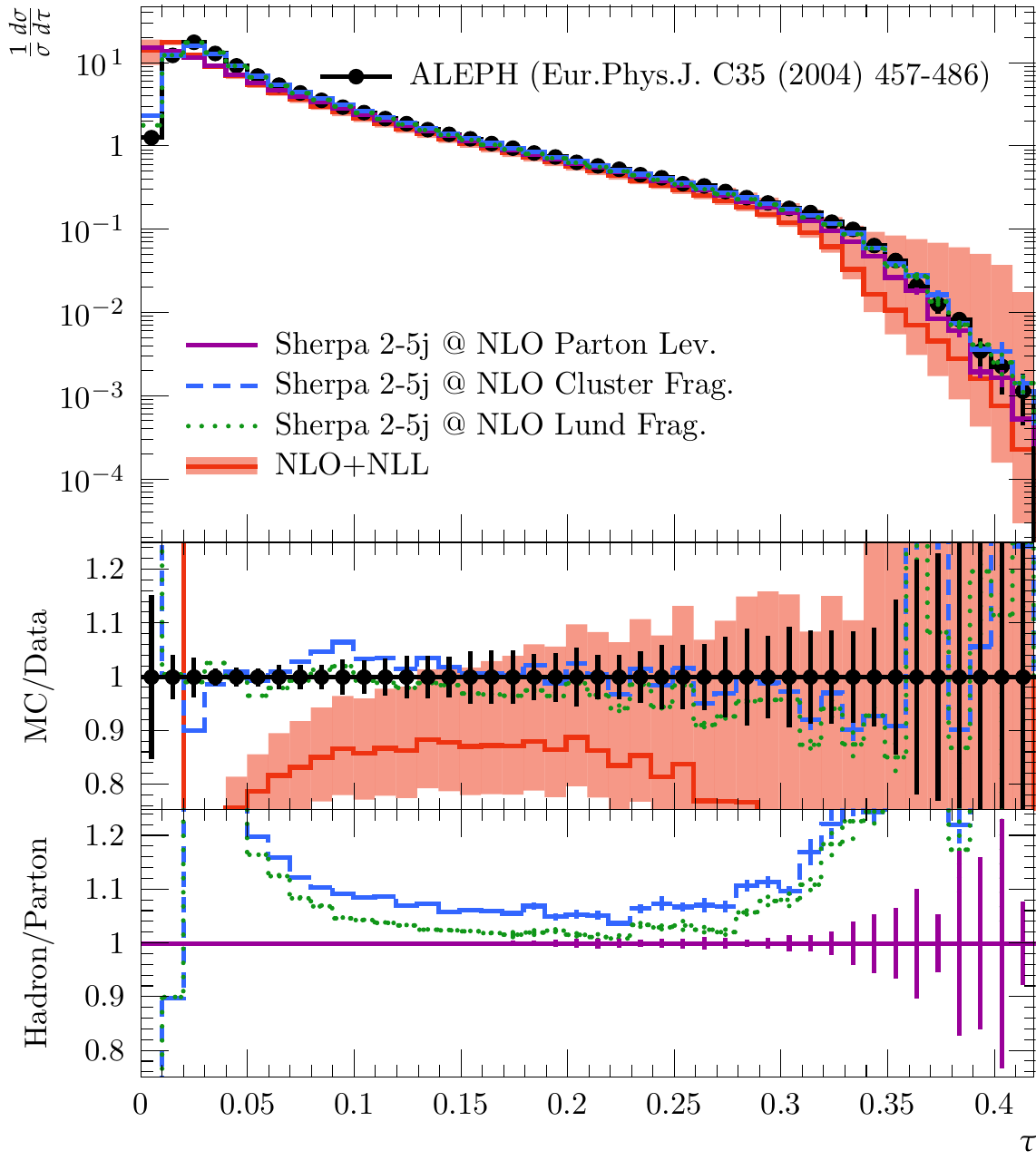}
  \caption{Plain thrust distribution as measured by ALEPH compared to \sherpa\ at
    parton level and with cluster and Lund fragmentation, and compared to the
    nominal resummed distribution matched to NLO. The lower panel shows the
    hadronisation corrections as obtained from \sherpa\ with the two fragmentation
    models considering MEPS@NLO with up to 5 jets at NLO. }\label{fig:plainThrust}
\end{figure}

For the generation of our pseudo data, used in the extractions of $\as$ later on, we consider the highest accuracy Monte Carlo sample which is generated using
NLO QCD matrix elements for $e^+ e^-\to2, 3, 4, 5$ partons. The MEPS@NLO merging parameter is set to
$y_{\textrm{cut}}=\left(Q_{\textrm{cut}}/E_{\textrm{CMS}}\right)^2=10^{-2}$. We evolve the strong coupling at
the two-loop order, assuming $\as(m_Z)=0.117$.\footnote{The \sherpa\ default value is
  $\as(m_Z)=0.118$. However, we observed a marginally better description of LEP1 observables,
  and in particular thrust, both for the cluster and the Lund string fragmentation
  using $\as(m_Z)=0.117$.} While for the cluster fragmentation model all parameters are
kept at their default values, we have set the main parameters of the Lund model to
\begin{equation}
  a = 0.3\;(\texttt{PARJ(41)}),\;\; b = 0.6\;\textrm{GeV}^{-2}\;(\texttt{PARJ(42)}),\;\; \sigma = 0.36\;\textrm{GeV}\;(\texttt{PARJ(21)})\,.
\end{equation}
Comparisons of \sherpa\ MEPS@NLO hadron-level predictions with LEP1 event-shape data have for
example been presented in~\cite{Gehrmann:2012yg}. To validate our event simulations, we present in
Fig.~\ref{fig:plainThrust} the plain thrust distribution as measured by ALEPH~\cite{Heister:2003aj}.
Shown there are the MEPS@NLO parton-level prediction, i.e. after parton showering, and hadron-level
results for the cluster and Lund fragmentation model. Furthermore, the purely perturbative
NLO+NLL resummed prediction with an estimate of the perturbative uncertainty, indicated by the red band,
cf. Sec.~\ref{sec:fits_uncertainties}, is given. The upper ratio plot compares theoretical predictions
with the experimental measurement. Apart from the first bin both hadron-level results are in good agreement
with data. However, the resummed calculation, without the inclusion of non-perturbative corrections,
significantly undershoots the data, in particular for $\tau<0.1$. 

The majority of this deviation originates from neglecting hadronisation effects in the analytic
calculation. This is apparent from the lower ratio plot. Here the two hadron-level results are compared with
the parton-shower-level prediction. For $\tau\approx 0.05$ the hadronisation corrections are of
order $20\%$. In order to justify the direct extraction of hadronisation corrections for our analytic
predictions of the soft-drop thrust observable we compiled Monte-Carlo simulations that better fit
the fixed-order accuracy of the matched resummed calculations. For this purpose we only consider
NLO QCD matrix elements for $e^+ e^-\to2, 3$ partons in our \sherpa\ MEPS@NLO simulations, with the
merging parameter still set to $y_{\textrm{cut}}=10^{-2}$. All other parameters are also kept unchanged.
A dedicated comparison of \sherpa\ parton-shower simulations and NLL accurate predictions of some
event-shape variables can be found in~\cite{Hoeche:2017jsi}. 

\begin{figure}
  \centering
  \includegraphics[width=\textwidth]{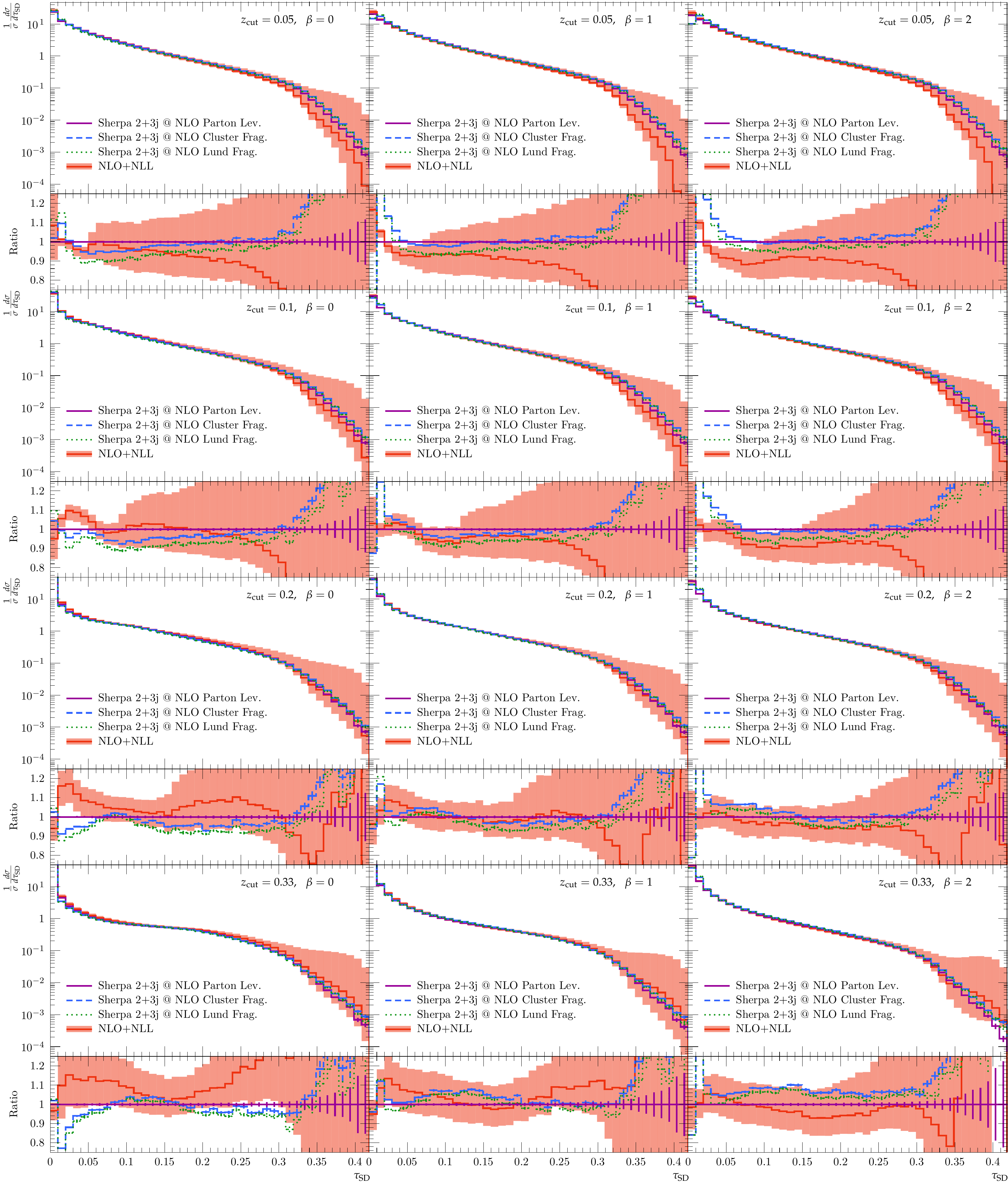}
  \caption{Comparison of NLO+NLL matched predictions for soft-drop thrust
    to \sherpa\ MEPS@NLO simulations with up to 3 jets at NLO at parton level
    and with cluster and string fragmentation applied.}\label{fig:Matching} 
\end{figure}

In Fig.~\ref{fig:Matching} we present results for soft-drop thrust, considering
$\zc=\{0.05,0.1,0.2,0.33\}$ (increasing from top to bottom) and $\beta=\{0,1,2\}$ (increasing from left to right).
Besides the \sherpa\ parton-level predictions the respective hadron-level results for cluster
and Lund fragmentation are given. Further, we show the NLO+NLL predictions. The lower panels
contain the respective ratios with the corresponding parton-level MC predictions.

We begin with noting that the hadronisation corrections for soft-drop thrust are indeed reduced
in comparison to plain thrust. As already seen in~\cite{Baron:2018nfz}, the region where the hadronisation
corrections are rather flat is extended towards smaller values of $\tau_{\textrm{SD}}$. For all values of
$\zc$ and $\beta$ the shape of the corrections from the cluster and Lund string model are very similar.
In particular for $\zc=0.33$ both hadron-level predictions agree very well also in their nominal size.
However, for the other $\zc$ values the differences remain in the few percent range. 

Besides the size of the MC hadronisation corrections to be used later in the
fits, Fig.~\ref{fig:Matching} contains the direct comparison of the NLO+NLL
calculations to the \sherpa\ parton-level prediction. For a wide range of the
observables the agreement is well within $\pm10\%$, the ratio between both
calculations being rather flat. Both results are certainly consistent within the
inherent uncertainties. Note that, in contrast to the analytic NLO+NLL prediction, the shower
result accounts for additional effects such as momentum conservation and finite
recoil~\cite{Hoeche:2017jsi}. In App.~\ref{sec:MC-val} further validation results
are provided. We compare \sherpa\ results against hadron-level predictions
from \pythia~\cite{Sjostrand:2007gs} and \herwig~\cite{Bahr:2008pv}. Furthermore,
we also compare to the \dire~\cite{Hoche:2015sya} algorithm for parton showering within
\sherpa~in conjunction with the default cluster model for
hadronisation.

\section{Hadronisation corrections}\label{sec:had_models}

Prior to describing our actual fits, we still need to address the issue of non-perturbative corrections from the
parton-to-hadron transition. As mentioned before, this is crucial for fits performed with the plain thrust distribution,
as perturbation theory alone is not able to reproduce the experimental data in the fitting range. We have seen that the
situation is greatly ameliorated if we employ the soft-drop version of thrust. However, although reduced in size,
hadronisation corrections play a non-negligible role also for soft-drop thrust. 

To supplement resummed and matched calculations with non-perturbative hadronisation corrections, two main approaches
can be found in the literature. On the one hand, analytical models of hadronisation can be constructed, based
on fairly general physical assumptions and depending on one or few input parameters only. On the other hand, one  can
exploit the hadronisation models used in Monte Carlo events generators to extract a numerical estimate of hadronisation
corrections by considering bin-by-bin ratios of the hadron-level and parton-level distribution. In the case of
plain thrust, analytic models were used, for instance,
in Refs.~\cite{Davison:2008vx,Abbate:2010xh,Abbate:2012jh,Gehrmann:2012sc,Hoang:2014wka},\footnote{Note that \cite{Abbate:2010xh,Abbate:2012jh,Hoang:2014wka} make use of a shape function approach including renormalon subtraction, which is more advanced than the approach presented in this work.} while examples of fits
of $\as$ exploiting Monte Carlo based hadronisation models can be found in
Refs.~\cite{Bethke:2008hf,Dissertori:2009ik,Dissertori:2009qa,OPAL:2011aa,Schieck:2012mp,Verbytskyi:2019zhh}.

\subsection{Monte-Carlo hadronisation model}
Analytical estimates of non-perturbative corrections in the presence
of grooming are characterised by additional complexities, which
require dedicated studies (see e.g.\ Ref.~\cite{Hoang:2019ceu} for a
recent study). Therefore, in the current analysis, we have
decided to resort to a Monte-Carlo based approach as our default model
for hadronisation corrections.
To this end the ratios of hadron- to parton-level
distributions are considered.  The hadronisation models implemented in general-purpose Monte Carlo event generators
depend on various parameters, whose values get tuned to data~\cite{Buckley:2011ms}. Hence, hadronisation corrections
obtained from Monte Carlo simulations inevitably include some perturbative contribution from missing higher-order
corrections that are specific to the underlying parton-level. Therefore, if we
are to supplement our analytic calculations with hadronisation corrections
extracted from Monte Carlo, a decent agreement with the parton-level Monte Carlo prediction should be in place. 
These comparisons were presented in Section~\ref{sec:MC} for the \sherpa~results at MEPS@NLO 2+3j level. 
There it is evident that the agreement is significantly improved for soft-drop thrust in contrast to plain thrust.
This makes the use of Monte-Carlo based hadronisation corrections more viable for soft-drop thrust at the current accuracy.

As the actual hadronisation corrections we take the hadron-to-parton ratios extracted from the \sherpa\ MEPS@NLO 2+3j
predictions, shown in Fig.~\ref{fig:Matching}. We consider the difference between the results for the cluster model and the Lund
string fragmentation as an estimate of the hadronisation related uncertainty. As illustrated in Appendix~\ref{sec:MC-val}
these two ratios provide a good coverage of the complete span found
for different Monte Carlo event generators. In addition, we
  include a comparison of our default method to a bin-by-bin migration-matrix approach and find only small differences between the two at our accuracy.

\subsection{Analytical hadronisation model}
As an alternative, we present the general concepts behind the analytical hadronisation model that was applied to ungroomed event (and jet)
shapes~\cite{Dasgupta:2007wa} and has been extended, more recently, to the case of mMDT~\cite{Dasgupta:2013ihk} and
soft drop~\cite{Marzani:2017kqd}. Details on the specific calculations underlying the results presented here are given in
Appendix~\ref{sec:hadr}.

In general we consider an additional (single) non-perturbative emission, that is supposed to be soft. For emissions below
an infrared factorisation scale $\mu_I$ the perturbative coupling is replaced by a universal, non-perturbatively
defined, finite quantity. When properly subtracting all perturbative contributions, one can thus estimate a hadronisation
contribution to the observable. For plain thrust this results in a shift of the observable value:
\begin{equation}
\delta\tau\(\Omega\)=2\frac{\Omega}{Q}\,,
\end{equation}
with $Q$ the centre of mass energy. The shift thereby is parametrised by the non-perturbative parameter
\begin{equation}
  \Omega \equiv C_F\Lambda(\mu_I) = C_F \int\limits_{0}^{\mu_I}\frac{dk_t}{\pi} \delta\as(k_t)
\end{equation}
that needs to be fitted simultaneously with $\as$.

For soft-drop thrust the situation is slightly more complicated. Typically, a non-perturbative emission will be soft
enough to be groomed away and thus has no impact on the observable value. However, this does not hold for a sufficiently
collinear emission, with a low enough transverse momentum to be called
non-perturbative but energetic enough to survive grooming.
These types of contributions are suppressed in $k_t$ and will depend on a separate non-perturbative parameter, that is a function
of $\beta$, cf. App.~\ref{sec:hadr_sd}. However, in the region we are interested in, the dominant hadronisation correction to soft-drop
thrust originates from a different configuration, namely a non-perturbative emission that survives grooming as it is protected
by hard emissions through the clustering history. Accordingly, the non-perturbative emission must lie within an angular cone
determined by the thrust value. The corresponding shift of the thrust distribution is calculated in the appendix and reads
\begin{equation}
\delta\tau\(\Omega\)=\tau^{1/2}\left<\left(z\left(1-z\right)\right)^{-1/2}\right>\frac{\Omega}{Q},
\end{equation}
where the average is calculated using as weight the appropriate QCD splitting function.
Note that the integral involved in the above average is sensitive to the transition point, as  is described in detail in App.~\ref{sec:hadr_sd}.

In addition to shifts in the value of thrust, there is a second way non-perturbative corrections affect soft-drop observables.
Specifically, after radiating a non-perturbative emission the energy
of one of the two hard subjets found by the soft-drop procedure ---
more precisely the softer one --- can be reduced so that it fails the
soft-drop condition and gets groomed away. This energy shift can be
taken into account in the form of a shift in the grooming parameter $\zc$:
\begin{equation}
\delta\zc\(\Omega\)=\frac{C_{A}}{C_F}\left(\frac{Z_{\text{SD}}\(1-Z_{\text{SD}}+\tau\)}{2\tau}\right)^{\beta/2}\frac{\(Z_{\text{SD}}-\tau\)\left(1-Z_{\text{SD}}\right)-\tau}{\sqrt{\tau \(Z_{\text{SD}}-\tau\)\left(1-Z_{\text{SD}}\right)}}\frac{\Omega}{Q}\,,
\end{equation}
where $Z_{\text{SD}}=z_{c}^{2/\left(2+\beta\right)}\left(2\tau\right)^{\beta/\left(2+\beta\right)}$. 
Because at LO it is not possible for an emission to be groomed away above transition point, we freeze this correction in that region.

The $\Omega$ parameter for both the groomed thrust shift and $\zc$ shift have the same corresponding integral definitions and approximations in their derivations, and are therefore assumed equal. While we use the same formal definition of the parameter $\Omega$ in
the case of plain and soft-drop thrust, different approximations were assumed in the derivation of the
respective hadronisation corrections and hence their numerical (fit) values are not expected to be identical.  
The model builds in the assumption that the observable and energy shifts are not too large and therefore the value of $\Omega$ is also assumed
rather small. Accordingly, we restrict our fits to  $\Omega \in [0,2]$ GeV.

\begin{figure}[h!]
	\centering
	\includegraphics[width=0.5\textwidth,page=1]{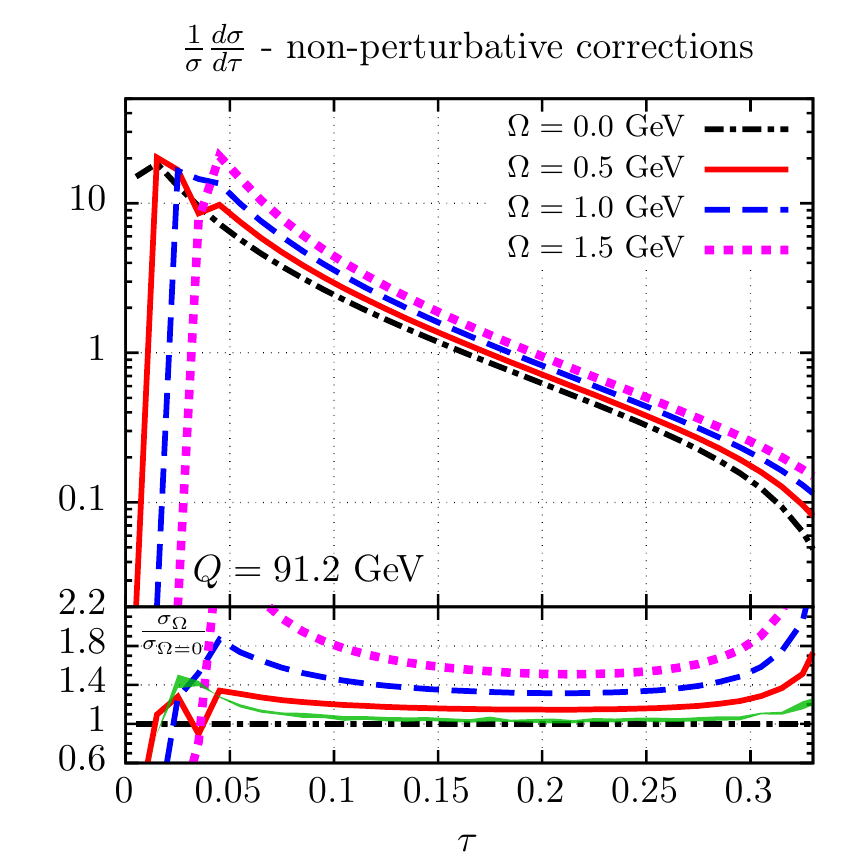}
	\caption{Comparison of the plain thrust NLO+NLL cross section under variation of the analytical hadronisation parameter $\Omega$.
          The lower panel displays the ratios with respect to the case of vanishing non-perturbative corrections, i.e. $\Omega=0$ GeV and the green band showing the hadron to parton level ratio for comparison.}\label{fig:Analytical_NP}
\end{figure}

\begin{figure}[h!]
\centering
\includegraphics[width=\textwidth]{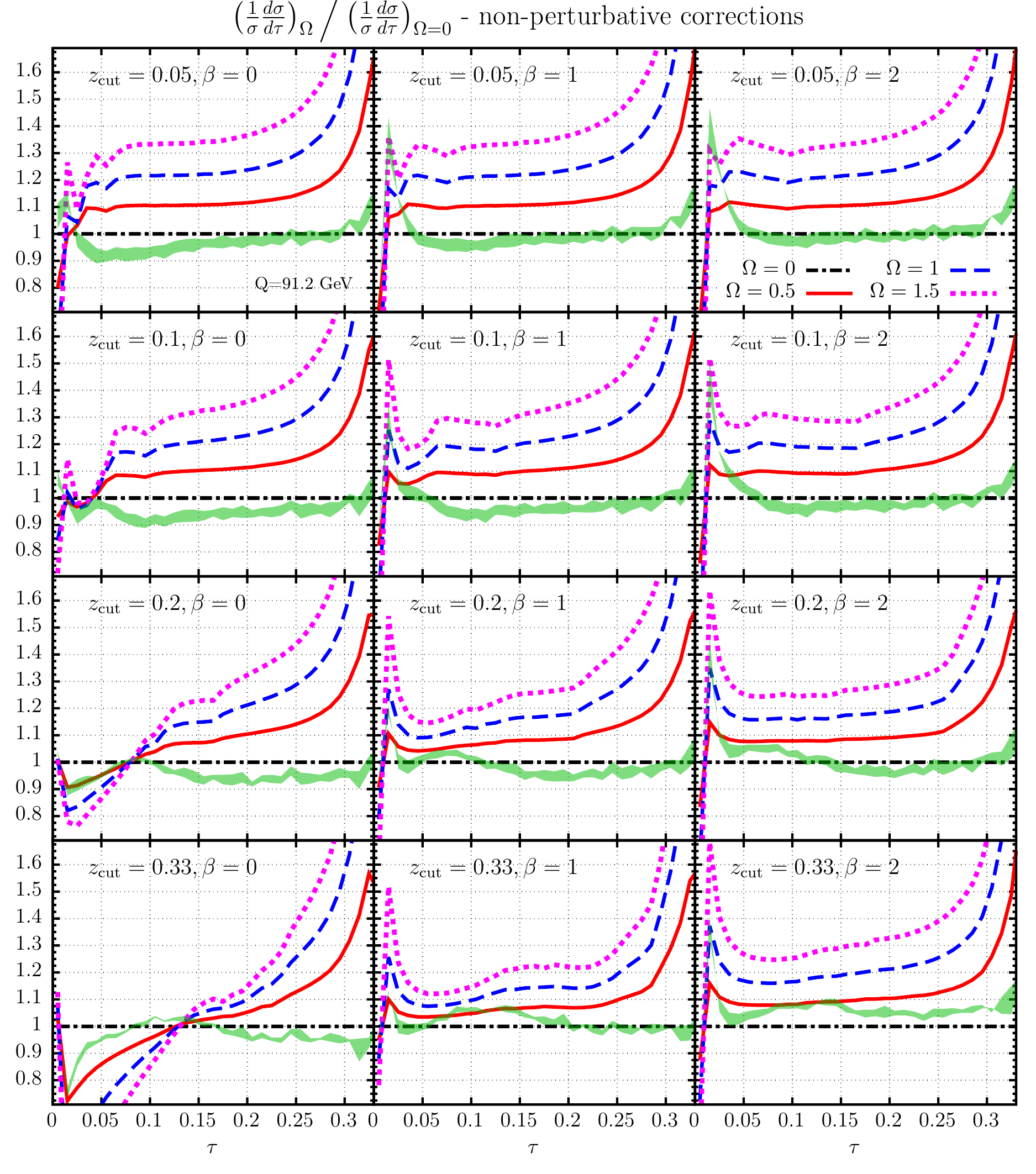}
\caption{Ratios for the soft-drop thrust NLO+NLL cross section under variation of the analytical hadronisation parameter $\Omega$ with respect to no non-perturbative corrections, i.e. $\Omega=0$ GeV. Where the green band shows the hadron to parton level ratio for comparison.}\label{fig:Analytical_NP_SD}
\end{figure}

The variation of plain thrust and soft-drop thrust under different assumptions for the non-perturbative parameter $\Omega$ is
illustrated in Figs.~\ref{fig:Analytical_NP} and~\ref{fig:Analytical_NP_SD}, respectively. For plain thrust the analytical hadronisation model
produces hadronisation corrections very similar in shape\footnote{
Note that, even if they are similar in shape, analytic and Monte-Carlo-based hadronisation corrections can differ significantly. For example, analytic hadronisation corrections can be made arbitrarily small by taking $\Omega\to 0$, while Monte-Carlo-based corrections are bounded around what is given by the different generators.}
to those extracted from Monte Carlo, cf. Fig.~\ref{fig:plainThrust}, though differences between parton level and resummation still cause different, but compatible, fitted values of $\as$.  
For equal values of
$\Omega$, the hadronisation contribution to soft-drop thrust is significantly reduced. However, the behaviour for the analytical model does not
fully replicate the hadron-to-parton ratios from Monte Carlo, cf. Fig.~\ref{fig:Matching}. For $\beta=0$ below the transition point ($\zc/2$)
the decrease in cross section corresponds to the behaviour seen in the Monte Carlo results. However, above the transition the $\zc$ shift has
little effect and therefore the thrust shift dominates. This leads to an increase in the cross section, which does not correspond to what is
seen in the Monte Carlo simulations.

Recently a more refined computation of non-perturbative corrections to the groomed jet mass for $e^+ e^-$ colliders was performed~\cite{Hoang:2019ceu}. This calculation can potentially be applied to the soft-drop thrust observable. However, as stressed in Ref.~\cite{Hoang:2019ceu}, the calculation is only applicable below the transition point and, in particular, it does not reduce to the standard, i.e.\ ungroomed, non-perturbative correction. Since the region neighbouring the transition point is quite relevant to the fit, further developments are needed to employ a field-theoretical description of hadronisation corrections in precision fits of the strong coupling.

\section{Fitting procedure}\label{sec:fit-as}
Having presented the theoretical calculations we are going to use for the determination of the strong coupling, as
well as the Monte Carlo tools employed to generate pseudo data and our approaches to account for hadronisation corrections,
we are now ready to discuss the actual fitting procedure.  The fits are based on a prescription similar to what has been done
for plain thrust, making use of a $\chi^2$ minimisation.

\subsection*{Fitting setup and uncertainty definition}\label{sec:fits_uncertainties}
The fit of $\as$ is performed by minimising the $\chi^2$ given by:
\begin{equation}
\chi^2=\sum_{i,j}\[\(\frac{1}{\sigma}\frac{d\sigma}{d\tau}\(\tau_i\)\)_{\rm exp}-\(\frac{1}{\sigma}\frac{d\sigma}{d\tau}\(\tau_i\)\)_{\rm th}\]V_{ij}^{-1}\[\(\frac{1}{\sigma}\frac{d\sigma}{d\tau}\(\tau_j\)\)_{\rm exp}-\(\frac{1}{\sigma}\frac{d\sigma}{d\tau}\(\tau_j\)\)_{\rm th}\]\,,
\end{equation}
where both the experimental and theoretical thrust distributions are
normalised to the respective inclusive cross section and the thrust distribution in a bin of width $d\tau=0.01$ is computed using differences of the cumulative distribution $\Sigma(\tau)$ at the edges of the bin. 
The sums in the definition of $\chi^2$ extend over the considered observable bins. The correlation matrix $V_{ij}$ contains the uncertainties based on the experimental data
\begin{equation}
V_{ij}=\delta_{ij}\sigma^2_{\rm stat}+\min\(\sigma^2_{{\rm sys},i},\sigma^2_{{\rm sys},j}\)
\end{equation}
as assumed for previous fits~\cite{Abbate:2010xh,Abbate:2012jh,Gehrmann:2012sc}. The uncertainties for our
\sherpa\ pseudo data are taken as directly proportional to the plain thrust ALEPH uncertainties:
\begin{eqnarray}
\sigma^{\rm MC}_{\rm stat}&=&\sigma^{\rm ALEPH}_{\rm stat}\sqrt{\frac{d\sigma^{\rm MC}/d\tau}{d\sigma^{\rm ALEPH}/d\tau}}\,,\\
\sigma^{\rm MC}_{\rm sys}&=&\sigma^{\rm ALEPH}_{\rm sys}\,.
\end{eqnarray}

The strong coupling constant is fitted for the central renormalisation scale choice $\mu_R/Q=1$.
The resummation scale, which appears in the logarithms of $\tau$ and
$\zc$ (cf.\ App.~\ref{sec:resummation}), is
also used at its central value $\mu_Q/Q \equiv x_L=1$. Finally multiplicative
matching is used and the power for the end-point correction is
$p=1$~\cite{Jones:2003yv} (cf.\ App.~\ref{sec:endpoint}). When the Monte Carlo based hadronisation model is
applied, the central fit is made using the average of the ratios of cluster
and Lund string fragmentation to the parton-level prediction.

The experimental uncertainty on the central results is determined by the range
of $\as$ values with $\Delta\chi^2=1$ around the central value while keeping the
hadronisation parameter $\Omega$ fixed. The theory uncertainty is determined by
varying simultaneously the theory inputs, i.e.\ the parameter $p$ assuming $p=1$
and $2$, the matching scheme by switching between the multiplicative and LogR
prescription and by 7-point variations of the perturbative scales
(excluding $\mu_R/Q=x_L=2$ and $\mu_R/Q=x_L=1/2$) using the best fit
for each of these variations. The corresponding uncertainty is then defined by
the difference between the central value fit and the found minimum and maximum
variations in the fitted value of $\as$ (and $\Omega$). 

The hadronisation uncertainty is model dependent. For the analytical model the
uncertainty is determined by $\Delta\chi^2=1$ as above while allowing for a
variation of $\Omega$. The previously mentioned experimental uncertainty needs
to be subtracted from this quadratically to prevent double counting. For the
Monte-Carlo based model the hadronisation uncertainty is determined by the range
between the best fit using either the cluster model or the Lund string
fragmentation.

\section{Results}\label{sec:results}
In this section we present our results for the extraction of $\as$ from fitting NLO+NLL predictions for the soft-drop
thrust observable to Monte-Carlo pseudo data. To account for the parton-to-hadron transition we employ both the analytical
approach and Monte-Carlo simulations. We are particularly interested in assessing the impact of soft-drop grooming on the stability
of the fits and their quality. To this end we compare results obtained for soft-drop thrust to those for plain thrust.

The default observable range used in the fits is $0.06\leq\tau\leq0.25$. The lower boundary equals the one used in the previous thrust
fits, cf.~\cite{Abbate:2010xh,Abbate:2012jh}.  However, the upper boundary is somewhat reduced, as in our study we work at a lower
fixed-order precision, i.e.\ NLO instead of NNLO QCD, resulting in a reduced accuracy for the distributions large-$\tau$ tail.

\begin{figure}[h!]
	\centering
	\begin{subfigure}[t]{.48\textwidth}
		\centering
		\includegraphics[width=\textwidth]{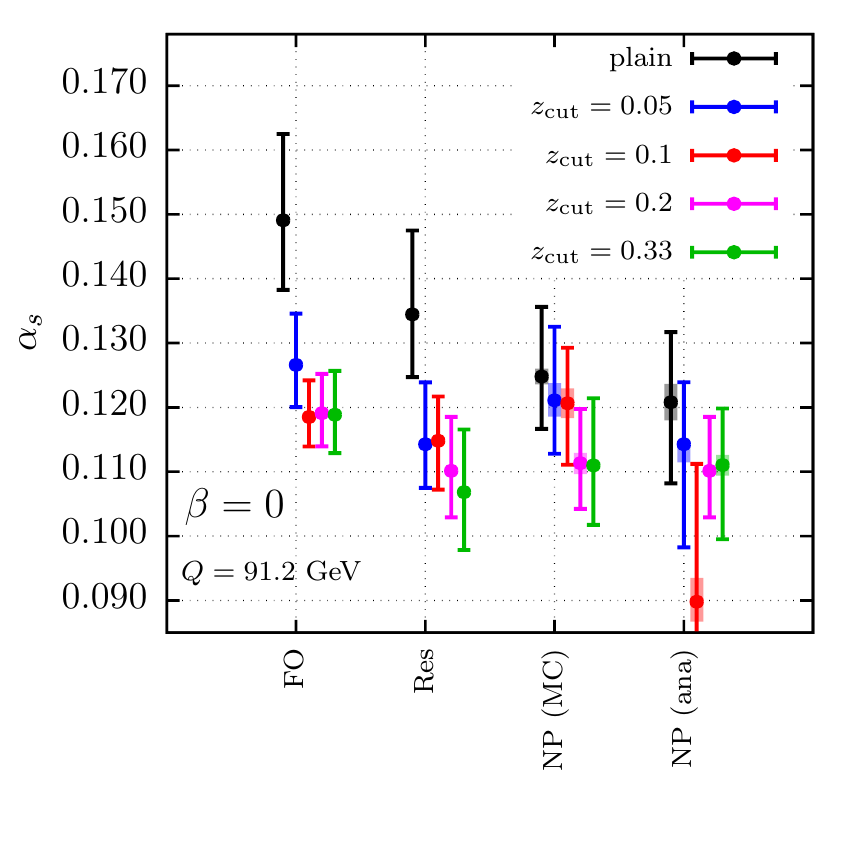}
	\end{subfigure}
	\hfill
	\begin{subfigure}[t]{.48\textwidth}
		\centering
		\includegraphics[width=\textwidth]{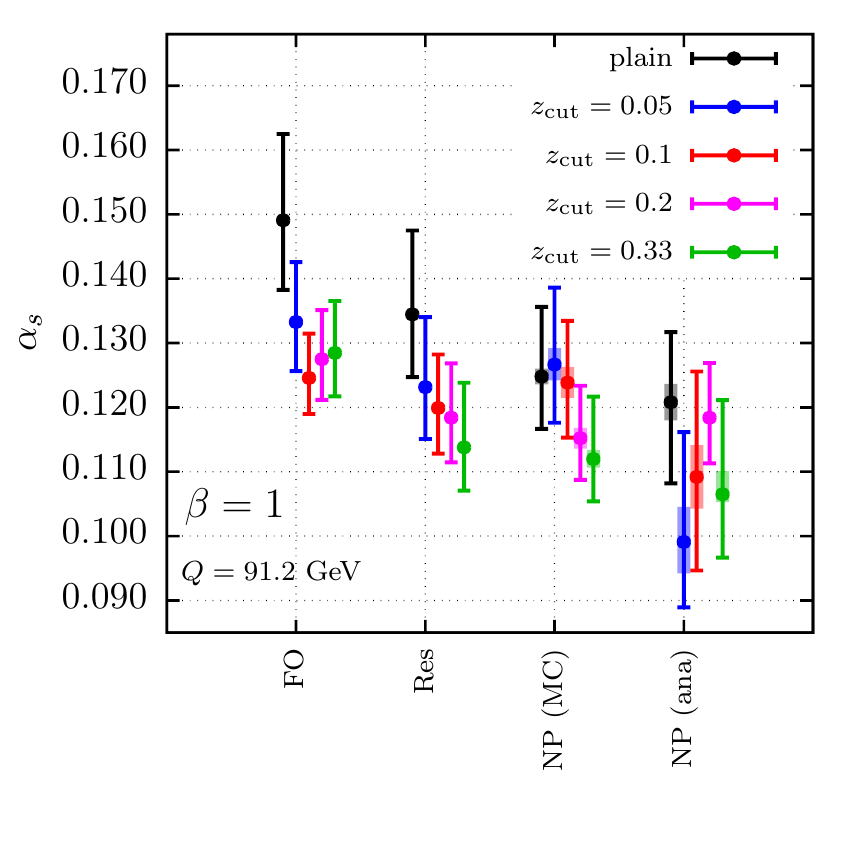}
	\end{subfigure}\\
	\begin{subfigure}[t]{.48\textwidth}
		\centering
		\includegraphics[width=\textwidth]{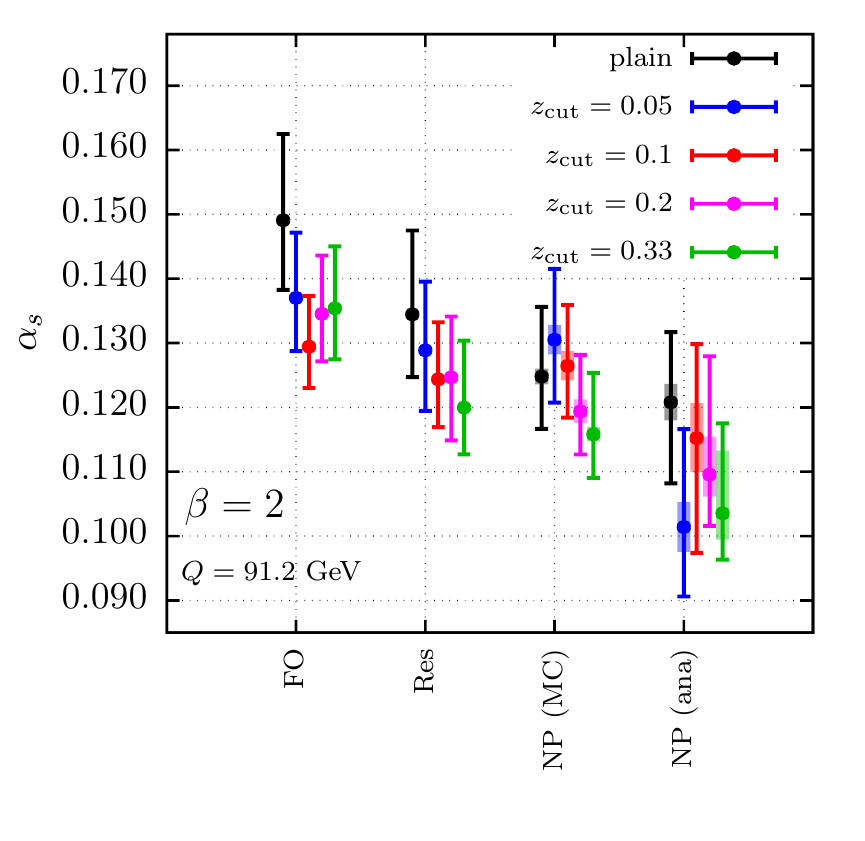}
	\end{subfigure}
	\caption{Results for fits of $\as(m_Z)$ to Monte-Carlo pseudo data using theoretical predictions
          at different levels: FO for NLO, Res for NLO+NLL and NP for the inclusion of non-perturbative effects.
          The non-perturbative effects are modelled based on either the Monte-Carlo based hadron-to-parton level
          ratios (NP (MC)) or an analytical model with a single parameter $\Omega$ (NP (ana)). The bands
          indicate the total uncertainty, the shaded region displays the hadronisation-related uncertainty.}\label{fig:as-fit}
\end{figure}

We begin by considering the stability of the extracted strong coupling under variations of the type of theoretical prediction entering
the fit. In Fig.~\ref{fig:as-fit} we present results for the best-fit $\as(m_Z)$ value and the associated total and hadronisation related
(shaded band) theoretical uncertainties. Besides the pure fixed-order result from \eventtwo~\cite{Catani:1996jh,Catani:1996vz} (FO),
we use the NLO+NLL resummed prediction matched to the NLO matrix element (Res), as well as this resummed result dressed with
non-perturbative corrections from the Monte-Carlo fragmentation approach (NP (MC)) and the analytic model (NP (ana)).
We present results for three values of the soft-drop angular exponent $\beta=0, 1, 2$.  For each value of $\beta$ we consider
$\zc=0.05, 0.1, 0.2, 0.33$, indicated by different colours. Furthermore, for comparison, we present the results obtained for plain thrust
in each plot.

From the plots we can draw some first general conclusion. The impact of incorporating NLL resummation is sizeable
both for the groomed and ungroomed observables. We observe a significant reduction of the best fit $\as(m_Z)$
in comparison to the fixed-order hypotheses. However, the role that non-perturbative corrections play is rather different.
The shift induced by the inclusion of hadronisation corrections is significantly reduced when soft-drop is employed.
This is true for both the analytical and the Monte-Carlo based models for hadronisation. Unfortunately, this reduced
impact of hadronisation corrections is not accompanied by a reduction of the associated uncertainties. 

However, while these general patterns appear to be well-established, there are noticeable counterexamples in particular
when using the analytical model for estimating hadronisation corrections. The first one standing out is $\beta=0$
and $\zc=0.1$ (shown in red in the upper left plot of Fig.~\ref{fig:as-fit}). For this particular set of soft-drop
parameters the lower boundary of the fitting region is just above the LO transition point. However, in the analytic
hadronisation model the transition point can, for large enough value of $\Omega$, be shifted into the fitting interval.
This effect leads to an odd fitting behaviour. It is dominant for $\beta=0$ as in this case transition-point effects
are more significant. Additionally, too little grooming such as $\zc=0.05$ and $\beta>0$ (shown in blue in the upper
right and lower plot of Fig.~\ref{fig:as-fit}) shows a significant decrease in the fitted value of $\as$, when the
analytic model of hadronisation is employed. An identical study for an independent resummation code with a different
resummation uncertainty treatment, which allows for a shift in the transition point, is presented in Appendix~\ref{sec:unc}.

  All in all, our study suggests that a tighter grooming  --- i.e.\ larger
values of $\zc$, namely $\zc=0.2$ and $\zc=0.33$, or smaller values of
$\beta$, namely $\beta=0$ or $\beta=1$ --- shows a significant
improvement compared to plain thrust with respect to both the shift in
the fitted value due to the inclusion of hadronisation effects and the
stability under variations of the hadronisation model. While the
reduced shift from the analytical hadronisation corrections is clearly
visible for the $\beta=0$ and $\beta=1$ cases, for $\beta=2$ this
shift is qualitatively similar to the one observed for plain thrust.
The results are further detailed in Table~\ref{tab:as-fits}. Besides the values obtained for $\as(m_Z)$ for the
various scenarios considered, we provide the best-fit $\chi^2$-values, and, for the case of the analytic hadronisation
model, the best-fit $\Omega$-parameter. For most of the groomed observables the $\chi^2/{\rm dof}$ for the NLO+NLL
matched calculation without any non-perturbative effects are already close to $1$. The only soft-drop parameter
combinations that yield larger $\chi^2$ values are the two cases that were pointed out previously, $\zc=0.1$
with $\beta=0$ and $\zc=0.05$ with $\beta=2$. An important validation of our approach is the consistency between
the value of $\as$ when fitting to ALEPH data instead of the pseudo-data for plain thrust. Additionally, it can
be seen that despite the reduced data in the fitting region due to grooming the experimental uncertainty does
not increase significantly, unless a significant amount of grooming is applied, for example $\zc=0.33$ and $\beta=0$.

%\begin{table}[t!]
%  \centering
  \afterpage{%
    \clearpage% Flush earlier float
    \thispagestyle{empty}% empty page style
    \begin{landscape}% Landscape page
      \centering
      \hspace*{-1.4cm}
      {\tabulinesep=1.2mm
	    \begin{tabu}{|c c|c c|c c c|c c|}
	      \hline
	      &   & Res &  & NP (ana) & & & NP (MC) & \\
	      $\zc$ & $\beta$  & $\as$(exp,th) & $\chi^2/{\rm dof}$ & $\as$(exp,had,th) & $\Omega$(exp,$\as$,th)[GeV] & $\chi^2/{\rm dof}$ & $\as$(exp,had,th) & $\chi^2/{\rm dof}$\\	        \hline
	     \multicolumn{2}{|c|}{plain} & $0.1345^{+0.0011+0.0130}_{-0.0011-0.0097}$ & $1.99$ & $0.1208^{+0.0010+0.0028+0.0105}_{-0.0010-0.0028-0.0122}$ & $0.347^{+0.025+0.069+0.234}_{-0.025-0.069-0.072}$ & $0.35$ & $0.1248^{+0.0010+0.0012+0.0107}_{-0.0010-0.0012-0.0080}$ & $1.38$ \\
	      \hline
	      0.05 & 0 & $0.1143^{+0.0011+0.0095}_{-0.0012-0.0067}$ & $0.85$ & $0.1143^{+0.0011+0.0000+0.0095}_{-0.0012-0.0028-0.0157}$ & $0.000^{+0.060+0.167+0.823}_{-0.000-0.000-0.000}$ & $0.90$ & $0.1211^{+0.0012+0.0026+0.0110}_{-0.0012-0.0025-0.0078}$ & $0.76$ \\
	      & 1 & $0.1231^{+0.0012+0.0108}_{-0.0012-0.0080}$ & $1.20$ & $0.0991^{+0.0012+0.0054+0.0162}_{-0.0012-0.0048-0.0089}$ & $1.567^{+0.077+0.312+0.433}_{-0.077-0.350-0.646}$ & $0.49$ & $0.1267^{+0.0012+0.0025+0.0116}_{-0.0012-0.0025-0.0086}$ & $1.66$ \\
	      & 2 & $0.1289^{+0.0013+0.0106}_{-0.0013-0.0093}$ & $4.34$ & $0.1014^{+0.0012+0.0039+0.0147}_{-0.0012-0.0039-0.0100}$ & $1.643^{+0.067+0.223+0.357}_{-0.067-0.222-0.533}$ & $1.59$ & $0.1305^{+0.0013+0.0023+0.0107}_{-0.0013-0.0023-0.0094}$ & $3.95$ \\
	      \hline
	      0.1 & 0 & $0.1148^{+0.0015+0.0067}_{-0.0015-0.0074}$ & $1.92$ & $0.0898^{+0.0014+0.0037+0.0211}_{-0.0014-0.0031-0.0045}$ & $1.773^{+0.097+0.237+0.227}_{-0.098-0.265-1.192}$ & $0.90$ & $0.1207^{+0.0016+0.0023+0.0082}_{-0.0016-0.0023-0.0092}$ & $2.81$ \\
	      & 1 & $0.1199^{+0.0014+0.0082}_{-0.0014-0.0070}$ & $0.64$ & $0.1092^{+0.0014+0.0049+0.0155}_{-0.0014-0.0049-0.0136}$ & $0.644^{+0.082+0.292+0.644}_{-0.083-0.285-0.498}$ & $0.64$ & $0.1239^{+0.0015+0.0024+0.0092}_{-0.0015-0.0024-0.0081}$ & $1.72$ \\
	      & 2 & $0.1244^{+0.0014+0.0088}_{-0.0014-0.0073}$ & $0.51$ & $0.1152^{+0.0014+0.0054+0.0135}_{-0.0014-0.0053-0.0170}$ & $0.515^{+0.076+0.295+0.791}_{-0.077-0.297-0.317}$ & $0.38$ & $0.1264^{+0.0014+0.0023+0.0091}_{-0.0014-0.0022-0.0076}$ & $1.10$ \\
	      \hline
	      0.2 & 0 & $0.1102^{+0.0022+0.0081}_{-0.0022-0.0069}$ & $0.74$ & $0.1102^{+0.0022+0.0000+0.0081}_{-0.0022-0.0000-0.0069}$ & $0.000^{+0.078+0.011+0.064}_{-0.000-0.000-0.000}$ & $0.78$ & $0.111^{+0.0022+0.0016+0.0080}_{-0.0022-0.0017-0.0065}$ & $0.79$ \\
	      & 1 & $0.1184^{+0.0018+0.0083}_{-0.0019-0.0067}$ & $0.77$ & $0.1184^{+0.0018+0.0000+0.0083}_{-0.0019-0.0000-0.0068}$ & $0.000^{+0.072+0.043+0.000}_{-0.000-0.000-0.000}$ & $0.82$ & $0.1152^{+0.0018+0.0016+0.0078}_{-0.0018-0.0016-0.0060}$ & $0.70$ \\
	      & 2 & $0.1247^{+0.0016+0.0093}_{-0.0016-0.0068}$ & $0.68$ & $0.1095^{+0.0015+0.0059+0.0174}_{-0.0015-0.0034-0.0070}$ & $1.180^{+0.007+0.000+0.424}_{-0.007-0.000-0.424}$ & $0.62$ & $0.1194^{+0.0016+0.0018+0.0085}_{-0.0016-0.0018-0.0062}$ & $0.89$ \\
	      \hline
	      0.33 & 0 & $0.1068^{+0.0034+0.0911}_{-0.0035-0.0083}$ & $1.46$ & $0.1110^{+0.0034+0.0016+0.0080}_{-0.0035-0.0016-0.0109}$ & $0.348^{+0.135+0.057+0.368}_{-0.137-0.069-0.191}$ & $1.26$ & $0.111^{+0.0035+0.0004+0.0098}_{-0.0036-0.0005-0.0085}$ & $1.11$ \\
	      & 1 & $0.1138^{+0.0022+0.0098}_{-0.0022-0.0063}$ & $1.21$ & $0.1065^{+0.0020+0.0036+0.0140}_{-0.0021-0.0012-0.0010}$ & $1.500^{+0.003+0.001+0.500}_{-0.003-0.001-0.936}$ & $1.05$ & $0.1120^{+0.0022+0.0014+0.0093}_{-0.0022-0.0014-0.0060}$ & $1.04$ \\
	      & 2 & $0.120^{+0.0017+0.0103}_{-0.0017-0.0071}$ & $0.89$ & $0.1035^{+0.0016+0.0097+0.0099}_{-0.0016-0.0041-0.0057}$ & $1.500^{+0.127+0.374+0.098}_{-0.147-0.887-0.385}$ & $0.78$ & $0.1158^{+0.0017+0.0011+0.0093}_{-0.0017-0.0011-0.0065}$ & $0.83$ \\
	       \hline
	      \multicolumn{2}{|c|}{plain (ALEPH)}  & $0.1331^{+0.0011+0.0126}_{-0.0011-0.0094}$ & $3.25$ & $0.1153^{+0.0010+0.0029+0.0095}_{-0.0010-0.0029-0.0103}$ & $0.454^{+0.024+0.070+0.191}_{-0.024-0.071-0.060}$ & $0.79$ & $0.1235^{+0.0010+0.0012+0.0105}_{-0.0010-0.0012-0.0077}$ & $1.00$ \\
	      	      \hline
	  \end{tabu}}
	  \captionof{table}{The results for the fits of $\as(m_Z)$ entering Fig.~\ref{fig:as-fit} with the associated values of $\Omega$ and $\chi^2/{\rm dof}$ values.
            In addition the breakdown of the different contributions to the uncertainties are included. All fits employ Monte-Carlo generated pseudo data.
            For comparison, in the last row, we also report the results of the fit when using ALEPH data instead.}\label{tab:as-fits}
    \end{landscape}
    \clearpage% Flush page
  }
%\end{table}

\begin{figure}[h!]
	\centering
	\begin{subfigure}[t]{.45\textwidth}
		\centering
		\includegraphics[width=\textwidth]{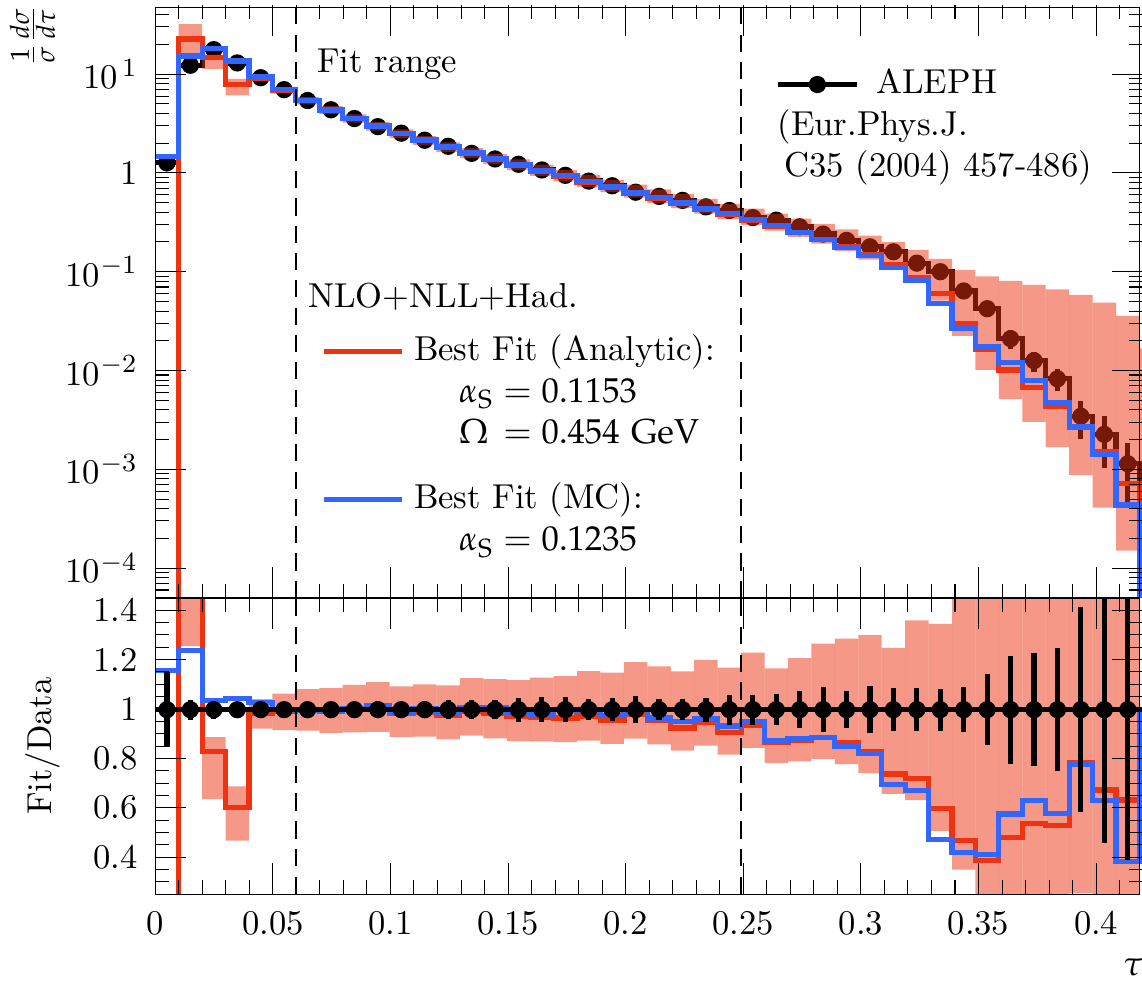}
	\end{subfigure}
	\hfill
	\begin{subfigure}[t]{.45\textwidth}
		\centering
		\includegraphics[width=\textwidth]{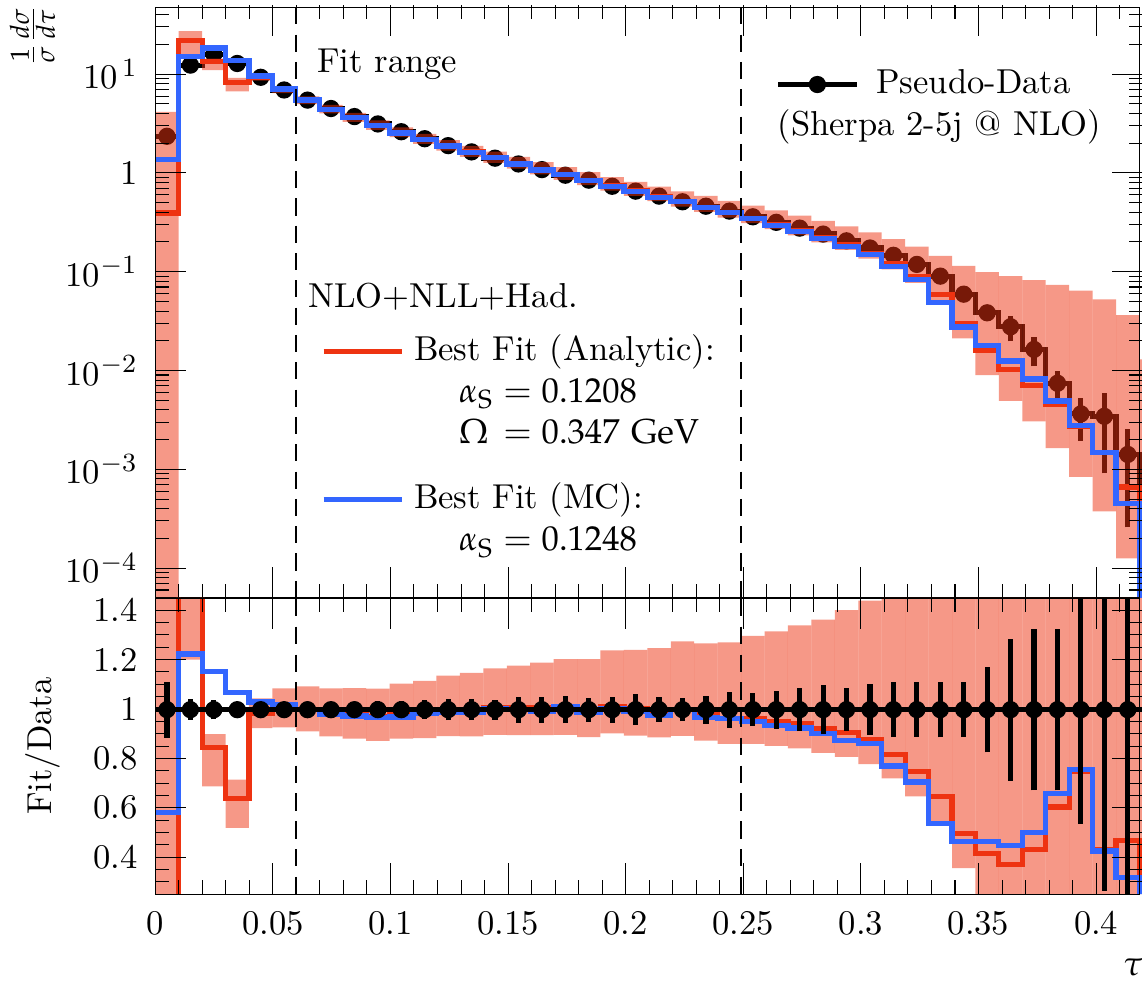}
	\end{subfigure}
	\caption{Plain thrust distribution as measured by ALEPH (left) and for Monte Carlo
          pseudo data (\sherpa\ MEPS@NLO 2-5j, right) compared to the NLO+NLL prediction with the
          best-fit values for $\as(m_Z)$ and the analytic (red) and Monte-Carlo (blue)
          hadronisation corrections. Uncertainties of the pseudo data are determined by
          rescaling of the uncertainties for the plain-thrust ALEPH data.}\label{fig:best_fit_plain}
\end{figure}

\begin{figure}[h!]
	\centering
	\includegraphics[width=\textwidth]{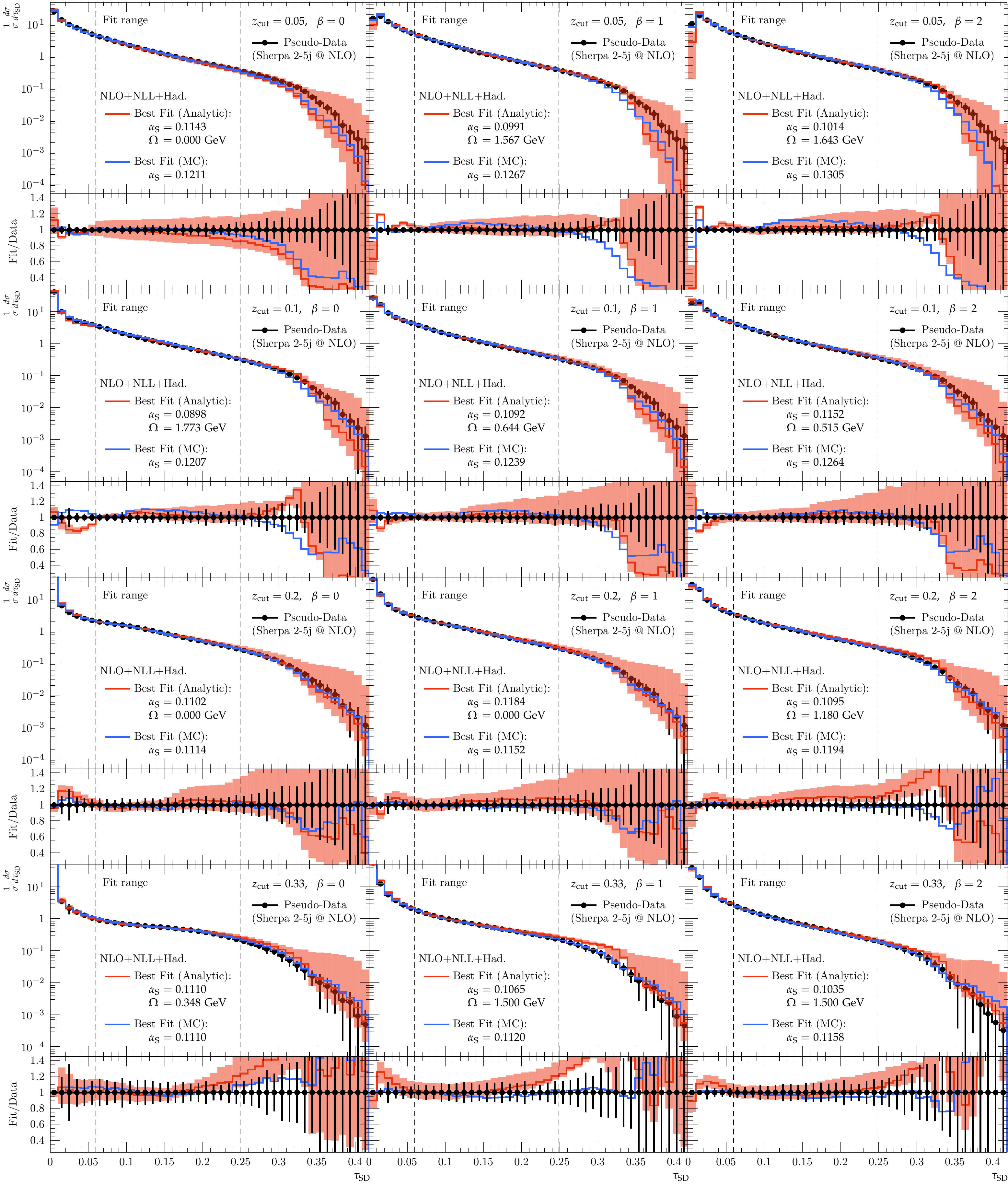}
	\caption{Pseudo-data (\sherpa\ MEPS@NLO 2-5j) and NLO+NLL result with best-fit
	  values for $\as(m_Z)$ and the analytic (red) and Monte-Carlo (blue) hadronisation
          corrections. Uncertainties of the pseudo data are determined by
          rescaling of the uncertainties for the ALEPH ungroomed thrust data.}\label{fig:best_fit_sd}
\end{figure}

To better visualise the results of our study, we can directly compare our theoretical predictions for the thrust observable
using the corresponding best-fit $\as$ with the (pseudo-)data. In Fig.~\ref{fig:best_fit_plain} we present this comparison
for plain thrust. The left plot shows the result for the fit to the
ALEPH data while the right plot shows those for the Monte-Carlo pseudo
data instead. Both plots illustrate a very good agreement of analytic predictions and both the actual collider and the pseudo
data over the whole fitting range. The red band represents the theoretical uncertainty on the fitted distribution
for the analytical model. The differences to the distribution with hadronisation
corrections based on the Monte Carlo model are small compared to the overall
size of the uncertainty, so we choose not to include a separate error band for this to
keep the plot clearer. It can also be seen that the behaviour in the small thrust
region would be hard to reproduce using the analytic model. The
corresponding results for soft-drop thrust, compared to our pseudo data, are presented in
Fig.~\ref{fig:best_fit_sd}, for all the $\zc$ and $\beta$ values considered in this study. Again, a good agreement is found.
For the considered fit range the biggest deviations are observed for large values of thrust when using the analytic
model. The agreement in this region could potentially be improved by including NNLO fixed-order corrections,
cf.~\cite{Kardos:2018kth}. When the Monte-Carlo based hadronisation model is used the most sizeable deviations are found
in the centre of the fitting range, which may also be resolved by
increasing the calculational accuracy.
One also sees that the deviations are
larger for the more problematic cases discussed above, $\zc=0.1$,
$\beta=0$ and $\zc=0.05$, $\beta=2$, while they remain very small for
$\zc=0.2$, $\beta=0,1$.  

All the results presented so far have been obtained performing the fits in the range that is typically employed
in $\as$ extractions using plain thrust, i.e.\ $0.06\leq\tau\leq0.25$. In particular, the lower bound of this interval
is a consequence of sizeable non-perturbative corrections for even smaller values of thrust, signalling the breakdown of
the perturbative approach. However, a key observation of
Ref.~\cite{Baron:2018nfz} was that the impact of non-perturbative
corrections on the soft-drop thrust distribution is consistently below 10\% for a much wider observable range compared to
plain thrust. Therefore, we can take advantage of this behaviour and push the lower bound of the fitting range to smaller
values of $\tau$, while retaining perturbativity. This is a key property of soft-drop observables, which has been exploited
also in the context of jet-mass measurements at the LHC~\cite{Marzani:2017mva,Marzani:2017kqd}.

\begin{figure}[h!]
	\centering
	\begin{subfigure}[t]{.4\textwidth}
		\centering
		\includegraphics[width=\textwidth]{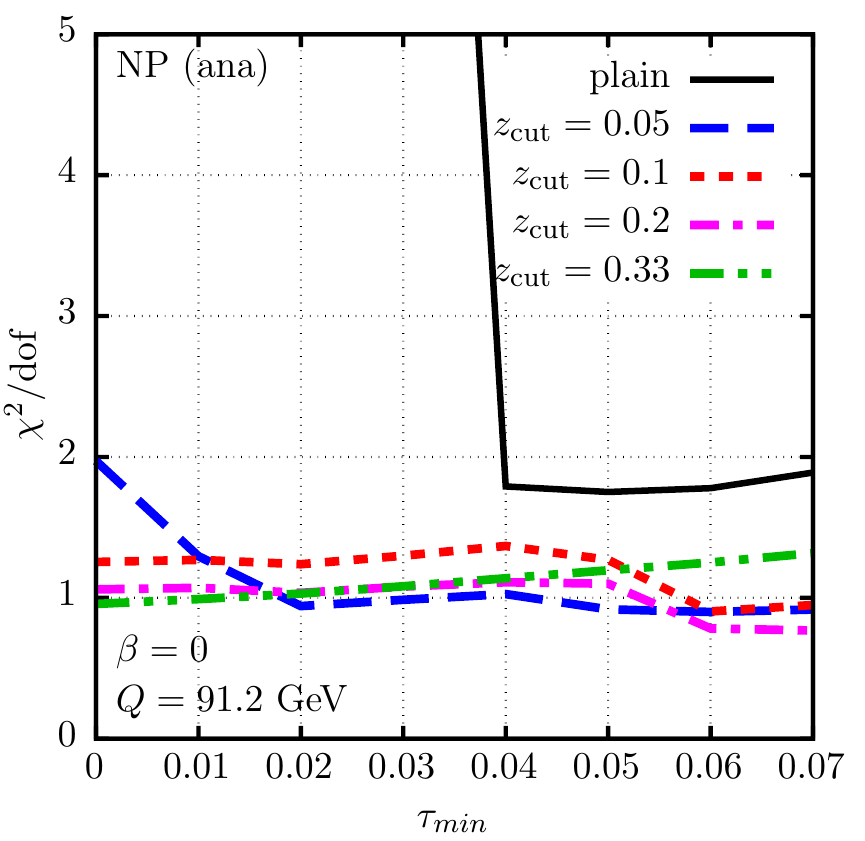}
	\end{subfigure}
	\hspace{0.7cm}
	\begin{subfigure}[t]{.4\textwidth}
		\centering
		\includegraphics[width=\textwidth]{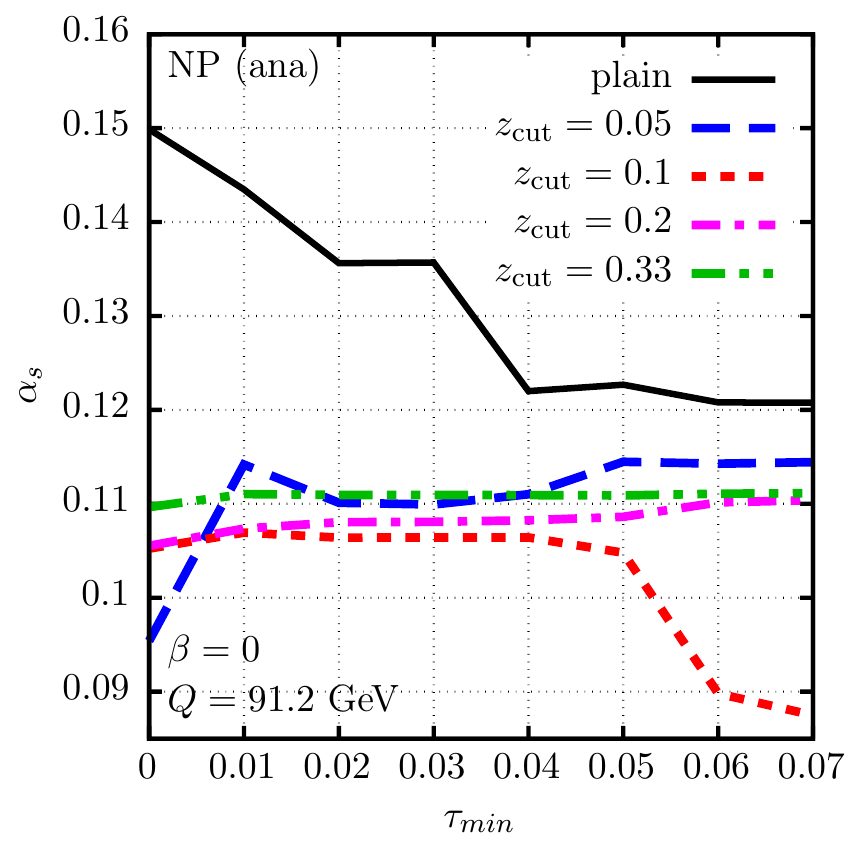}
	\end{subfigure}\\
	\begin{subfigure}[t]{.4\textwidth}
		\centering
		\includegraphics[width=\textwidth]{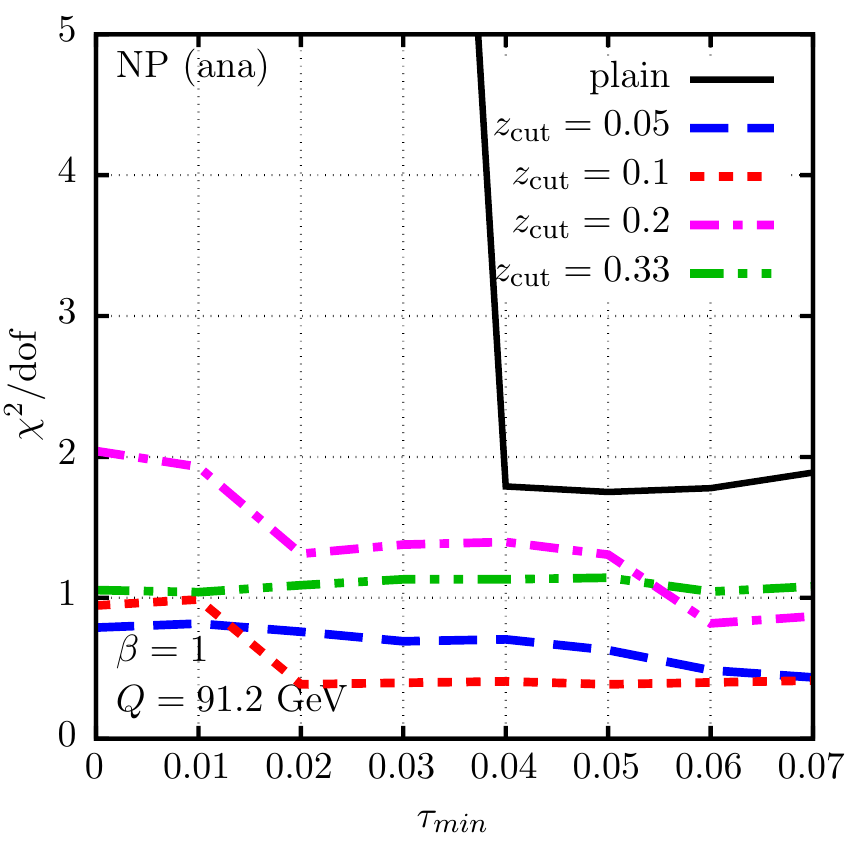}
	\end{subfigure}
	\hspace{0.7cm}
	\begin{subfigure}[t]{.4\textwidth}
		\centering
		\includegraphics[width=\textwidth]{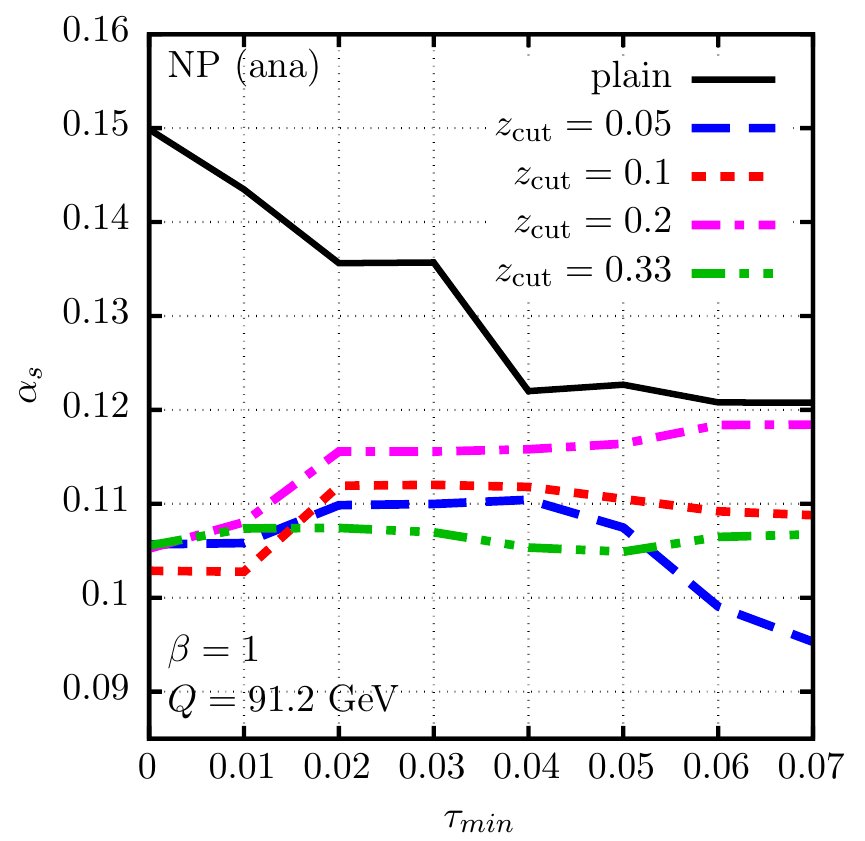}
	\end{subfigure}\\
	\begin{subfigure}[t]{.4\textwidth}
		\centering
		\includegraphics[width=\textwidth]{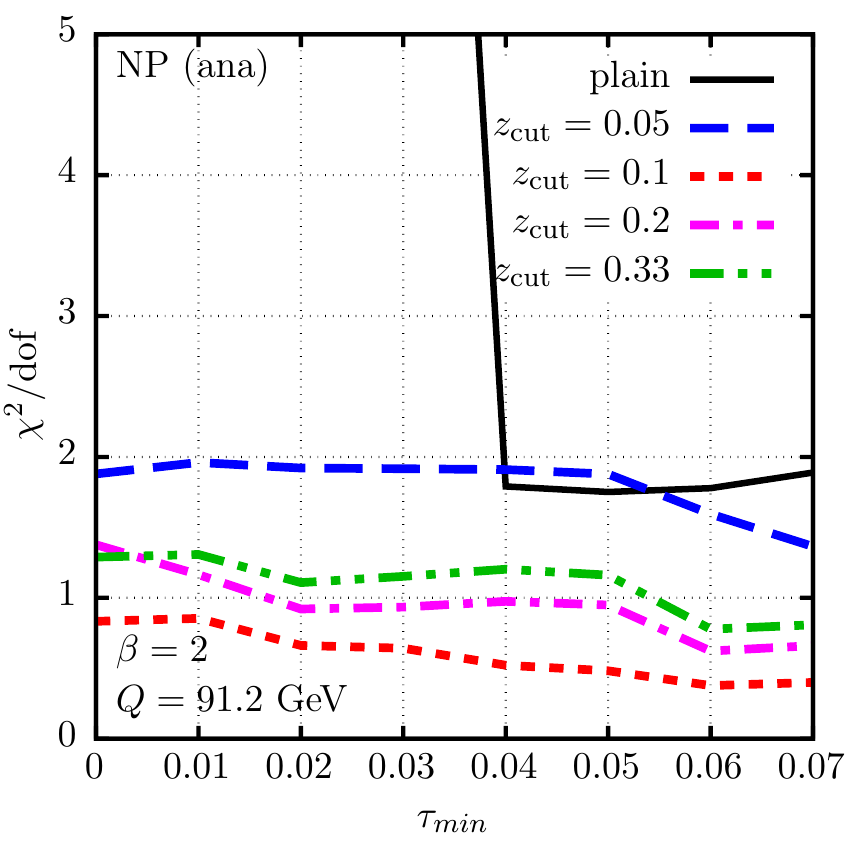}
	\end{subfigure}
	\hspace{0.7cm}
	\begin{subfigure}[t]{.4\textwidth}
		\centering
		\includegraphics[width=\textwidth]{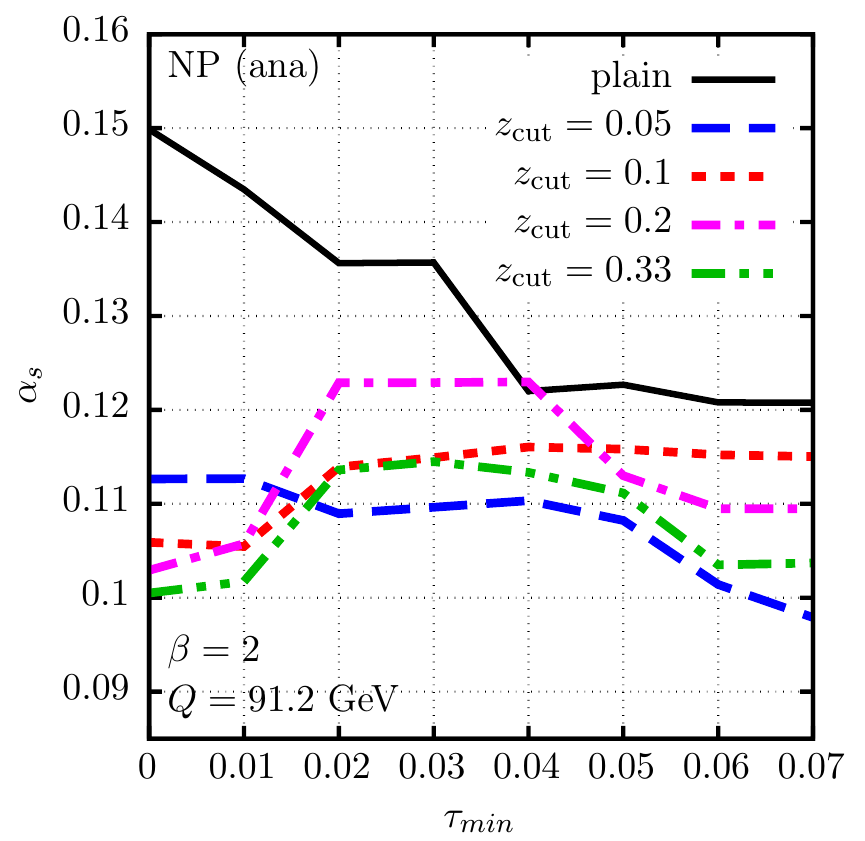}
	\end{subfigure}
	\caption{The best-fit $\chi^2/dof$ (left column) and $\as(m_Z)$ (right column) values
          as a function of the lower bound of the fitting range, $\tau_\mathrm{min}$. Shown are results
          for plain thrust and for soft-drop thrust, using different combinations of the
          soft-drop parameters $\beta$ and $\zc$.}\label{fig:Range_variation}
\end{figure}

In order to quantitatively assess this observation, we perform several fits for the strong coupling reducing the lower
bound of the fitting region. The results are reported in Fig.~\ref{fig:Range_variation}. As a measure of the fit quality,
we present in the left column the resulting $\chi^2/{\rm dof}$ values as a function of the lower bound of the fitting range
$\tau_\mathrm{min}$, for plain thrust and the various soft-drop parameters. The fits are performed with the analytic model for
the hadronisation corrections. It is apparent that the fit quality for plain thrust rapidly deteriorates below
$\tau_\mathrm{min}\simeq0.04$. Non-perturbative corrections become so large that they invalidate not only the perturbative
approach but also the assumptions that go into the rather simple analytic model of hadronisation. Indeed the deterioration
of the $\chi^2/{\rm dof}$ is not as dramatic, when using the Monte-Carlo based model for hadronisation. However, the fit
quality as measured by the $\chi^2/{\rm dof}$ for soft-drop thrust depends very weakly on the value of $\tau_\mathrm{min}$ only.
It is consistently better than what is obtained for plain thrust. 

Similar conclusions can be drawn by looking at the right column of Fig.~\ref{fig:Range_variation}, where the best-fit
$\as(m_Z)$ values are shown as a function of the lower bound of the fitting region $\tau_\mathrm{min}$. The best-fit $\as$ obtained
with plain thrust exhibits a significant dependence on $\tau_\mathrm{min}$, while the values obtained with soft-drop thrust are
rather stable under variation of $\tau_\mathrm{min}$. Furthermore, we remind the reader that in Fig.~\ref{fig:as-fit} we observed
an issue for the fit performed with  $\zc=0.1$ with $\beta=0$, resulting in a rather small value of the strong coupling
(with large uncertainties). We argued that this originates from the proximity of the transition point and the lower bound
of the fitting range, causing an instability in the modelling. The top right plot in Fig.~\ref{fig:Range_variation}
fully supports this interpretation and indeed shows that the issue can be resolved by pushing the lower bound of the
fitting range to smaller values. Choosing $\tau_\mathrm{min}\lesssim0.05$, we obtain fitted values of $\as$ that are compatible
with all the other combinations of soft-drop parameters.

In addition to the fitted value of $\as\(m_Z\)$ and the value of $\chi^2$ one can investigate the dependence of the
uncertainty on the increased fitting range. However, at the current theoretical precision the impact on the uncertainty
is found to be not significant. 

\section{Conclusion}\label{sec:conclusion}
In previous work it has been shown that the soft-drop thrust observable features reduced sensitivity to non-perturbative effects.
Motivated by this observation we have considered fits of the strong-coupling constant for this variable. We have analysed the
impact of both resummation and hadronisation corrections and compared to plain thrust. Two different approaches to estimate
hadronisation corrections have been used, one based on Monte-Carlo results from \sherpa~and another based on analytical computations. 

The Monte-Carlo based model turns out more viable for soft-drop thrust than for plain thrust. This roots in an improved
agreement of parton-level predictions based on NLO QCD matrix-element plus parton-shower simulations and the NLO+NLL analytical
results. Through the use of this model we have shown a significant decrease in the shift of the extracted $\as(m_Z)$ due to
hadronisation effects.

As an alternative an analytical hadronisation model for soft-drop thrust has been presented. Also here the general trend shows
an improvement over plain thrust, however, some outliers have been noted. In particular lower values of $\zc$ show issues in
the fits related to the transition point. In addition, for the analytical model we have shown that it is viable to extend the
fitting range to significantly lower values of the observable, which is not possible for plain thrust when using the
analytical model. Despite these improvements for the analytic
hadronisation model it still shows a relatively different
behaviour compared to the Monte-Carlo based approach. This indicates the need for a more detailed analysis of the analytical model
based on a more elaborate computation.

These conclusions were all made based on an analytical computation at NLO+NLL accuracy. However, for a future accurate measurement
of the strong coupling this will need to be extended to at least NNLO+NNLL precision. The NNLO computation presented
in~\cite{Kardos:2018kth} has been performed for the original definition of soft-drop thrust and could be extended to the
collinear-safe definition used in this work. The NNLL accuracy has been achieved in the limit $\tau\ll\zc\ll 1$, however
in the region relevant to this fit transition-point effects ($\tau\sim\zc$) will also need to be taken into account up to
NNLL. Finally, due to the higher values of $\zc$ used in this analysis
the impact of finite $\zc$ corrections (in particular for $\beta=0$) needs
to be studied. 

Finally, we note that, in order to reduce the impact of
transition-point corrections and to simplify their computation, it
would be worth considering other observables and modified
definitions of the soft-drop grooming procedure itself. Similar studies can also be performed for event shapes at the LHC,
where, besides hadronisation, in particular the underlying event obscures the comparison of perturbative predictions with
actual collider data. 
In conclusion, the work presented here shows promising possibilities for the application of soft drop to the measurement
of the strong-coupling constant. However, additional work will need to go into improving the theoretical accuracy to result
in a precision extraction entering the world average.

\begin{acknowledgments}
  We thank Pier Monni for the useful discussions.
  The work leading to this publication was supported by the German Academic Exchange Service (DAAD) with funds from the German
  Federal Ministry of Education and Research (BMBF) and the People Programme (Marie Curie Actions) of the European Union Seventh
  Framework Programme (FP7/2007-2013) under REA grant agreement n.\ 605728 (P.R.I.M.E.\ Postdoctoral Researchers International
  Mobility Experience). This work is also supported by the curiosity-driven grant "Using jets to challenge the Standard Model
  of particle physics" from Universit\`a di Genova. SS acknowledges funding from the European Union's Horizon 2020 research
  and innovation programme as part of the Marie Sk\l{}odowska-Curie Innovative Training Network MCnetITN3
  (grant agreement no. 722104), the Fulbright-Cottrell Award and from BMBF (contract 05H18MGCA1). DR acknowledges support
  from the German-American Fulbright Commission allowing him to stay at Fermi
  National Accelerator Laboratory (Fermilab), a U.S. Department of Energy,
  Office of Science, HEP User Facility. Fermilab is managed by Fermi Research
  Alliance, LLC (FRA), acting under Contract No.\ DE--AC02--07CH11359.
  VT also thanks the CNRS and the IPhT for financial support and
  hospitality during the early stages of this work.
  GS\ is supported in part by the French Agence Nationale de la
  Recherche, under grant ANR-15-CE31-0016.
\end{acknowledgments}

\appendix

\section{Collinear safety}\label{sec:col-safe}

In this Appendix we show that the rescaling factor in front of
Eq.~(\ref{eq:sd-thrust}) is crucial to make soft-drop thrust
infrared-and-collinear safe for the modified Mass-Drop tagger (i.e.\
soft-drop with $\beta=0$).

Consider the situation where there is a single massless particle, of
energy $E_R$ in the right hemisphere. The square bracket in
Eq~(\ref{eq:sd-thrust}) is then
\begin{equation}\label{eq:sd-thrust-1R}
\tau_\text{SD}^\text{(unscaled)} = 1-\frac{E_R+P_L}{E_R+E_L} = \frac{E_L-P_L}{E_R+E_L},
\end{equation}
with
$P_L=\sum_{i\in\mathcal{H}^{L}_{\text{SD}}}\left|\vec{n}_L\cdot\vec{p}_i\right|$
and $E_L=\sum_{i\in\mathcal{H}^{L}_{\text{SD}}}\left|\vec{p}_i\right|$
the contributions of the left hemisphere to the numerator and
denominator respectively (after soft-drop).

If the particle in the right hemisphere splits collinearly in two
particles of energies $z E_R$ and $(1-z)E_R$ the softest of these two
particles, say the one with energy $zE_R$, will be groomed away and
the unscaled soft-drop thrust becomes
\begin{equation}\label{eq:sd-thrust-1R+1coll}
  \tau_\text{SD, coll}^\text{(unscaled)}
  = 1-\frac{(1-z)E_R+P_L}{(1-z)E_R+E_L}
  = \frac{E_L-P_L}{(1-z)E_R+E_L}\,,
\end{equation}
which obviously differs from~(\ref{eq:sd-thrust-1R}) by a finite
amount independent of the angle of the collinear emission in the right
hemisphere.
For the corresponding virtual corrections (unscaled) soft-drop thrust
would still be given by~(\ref{eq:sd-thrust-1R}), yielding a
mis-cancellation between real and virtual contributions and hence a
collinear unsafety.
It is worth noting that this collinear divergence starts at
${\cal {O}}(\as^2)$ since at ${\cal {O}}(\as)$ there is only
one particle in the left hemisphere (the only emission being the
collinear one) and $E_L=P_L$.

\begin{figure}
  \centering
  \includegraphics[width=0.5\textwidth]{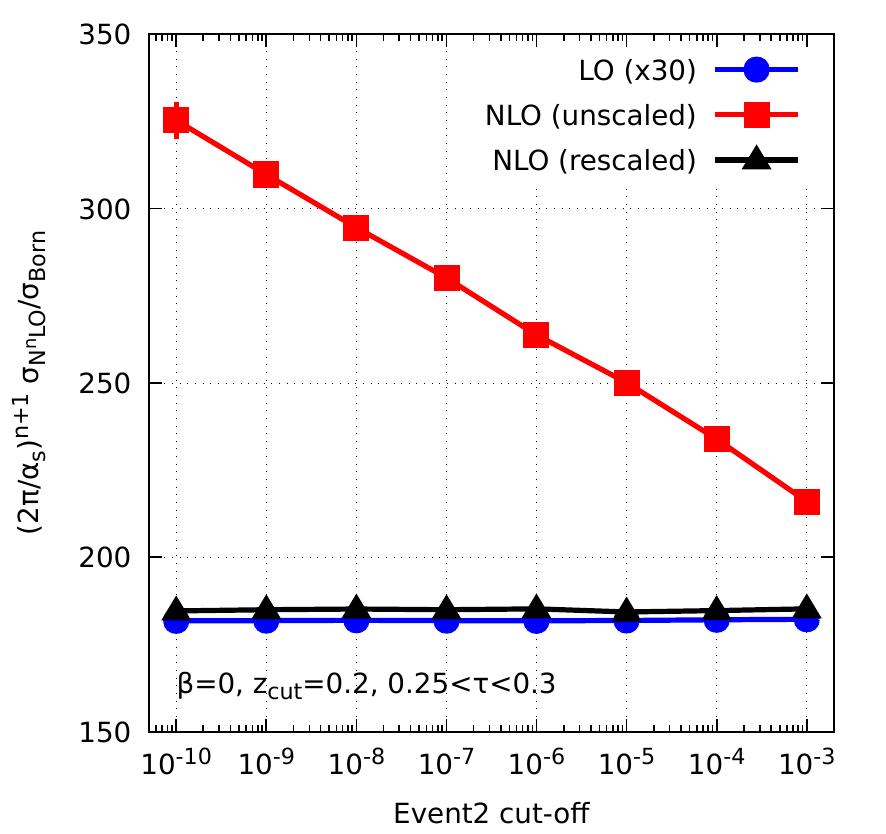}
  \caption{Illustration of the collinear divergence for unscaled
    soft-drop thrust using \eventtwo\
    simulations.}\label{fig:beta0-coll-div}
\end{figure}

If instead one introduces the scaling factor of
Eq.~(\ref{eq:sd-thrust}), one gets
\begin{equation}
  \tau_\text{SD}  =  \tau_\text{SD, coll} = \frac{E_L-P_L}{Q},
\end{equation}
and collinear safety is restored.

To better illustrate the divergence, Fig.~\ref{fig:beta0-coll-div}
shows the LO (${\cal {O}}(\as)$) and NLO
(${\cal {O}}(\as^2)$) cross sections for soft-drop thrust, with
$\beta=0$ and $z_\text{cut}=0.2$, to be between 0.25 and 0.3 as
obtained using \eventtwo~\cite{Catani:1996jh,Catani:1996vz}. The
results are presented as a function of the soft cut-off used in the
program. It is evident from the figure that the unscaled version of
soft-drop thrust exhibits a logarithmic divergence as the cut-off is
decreased while the rescaled version is collinear safe.

We note that this collinear unsafety issue is only present for
$\beta=0$. Indeed, when $\beta>0$ the maximum value of $z$ depends on
the angle of the emissions and goes to zero like a power of this
angle. As a consequence, Eq.~(\ref{eq:sd-thrust-1R+1coll}) coincides
with~(\ref{eq:sd-thrust-1R}) in the collinear limit and the unsafety
is absent.

\section{Changes to the analytical calculation}\label{sec:mult-trans}
The analytical calculation we employ in this study does not deviate significantly
from what was previously presented in Ref.~\cite{Baron:2018nfz}. Therefore, in this
appendix we limit ourselves to the presentation of those expressions we actually
use and focus on any difference to the aforementioned paper, the notation of which
we follow rather closely.

\subsection{Resummation equations}\label{sec:resummation}
The all-order expression for the soft-drop thrust distribution, in the region where
the soft-drop condition is active can be written as~\cite{Baron:2018nfz}:
\begin{align}\label{eq:Integrated-xsec}
\Sigma\(\tau\)&=\frac{1}{2\pi i}\int_C \frac{dN}{N} \[1+\sum_{n=1}^{\infty}\(\frac{\as}{\pi}\)^{n}\tilde{C}^{\(n\)}\]e^{-R\(\lambda_{\bar{N}},\lambda_{\zc}\)}\,,
\end{align}
where $\tilde{C}$ encapsulates the constant contributions in $\tau$ and $\zc$, which are neglected at NLL accuracy,
and $\as$ is evaluated at scale $\mu$. The resummed exponent $R$ in
the conjugate moment space is given by
\begin{equation}
R\(\lambda_{\bar{N}},\lambda_{\zc}\) =
-\frac{1}{\as}f_1\(\lambda_{\bar{N}},\lambda_{\zc}\)-f_2\(\lambda_{\bar{N}},\lambda_{\zc}\)-\as
f_3\(\lambda_{\bar{N}},\lambda_{\zc}\) - \dots,
\end{equation}
with $\lambda_{\bar{N}}=\as b_0\log\(x_L\bar{N}\)$, where $\bar N= N e^{\gamma_E}$, and
$\lambda_{\zc}=\as b_0\log\(x_L 2^{\beta/2} \zc\)$. The parameter $x_L$ is an arbitrary rescaling factor
that we vary in order to estimate missing higher-order corrections in the resummation, cf. Sec.~\ref{sec:fits_uncertainties}.
Note, the term $f_3$ does not contribute at NLL accuracy. The remaining functions $f_i$ ($i=1,2$) can be
expressed through
\begin{eqnarray}
f_1^{K}\(\lambda_T\)&=&\frac{\Gamma_K^{(0)}}{2 b_0^2\pi}\[\(1+2\lambda_T\)\log\(1+2\lambda_T\)-2\lambda_T\],\\
f_2^{K}\(\lambda_T\)&=&\frac{\Gamma_K^{(1)}}{ b_0^2\pi^2}\[\lambda_T-\frac{1}{2}\log\(1+2\lambda_T\)\]+\frac{\Gamma_K^{(0)}b_1}{ 4b_0^3\pi}\[\log\(1+2\lambda_T\)\(2+\log\(1+2\lambda_T\)\)-4\lambda_T\]\nonumber\\
&&+\frac{\Gamma_K^{(0)}}{ 2b_0\pi}\log\(1+2\lambda_T\)L_\mu^K-\frac{\gamma_K^{(0)}}{2b_0\pi}\log\(1+2\lambda_T\),
\end{eqnarray}
according to 
\begin{eqnarray}
f_i\(x,y\)%&=&f_i^{S_G}\(p^{(\bar{N})}_{S_G}x+p^{(\zc)}_{S_G}y\)+2f_i^{S_C}\(p^{(\bar{N})}_{S_C}x+p^{(\zc)}_{S_C}y\)+2f_i^{J}\(p^{(\bar{N})}_{J}x+p^{(\zc)}_{J}y\)\nonumber\\
&=&f_i^{S_G}\(p^{(\zc)}_{S_G}y\)+2f_i^{S_C}\(p^{(\bar{N})}_{S_C}x+p^{(\zc)}_{S_C}y\)+2f_i^{J}\(p^{(\bar{N})}_{J}x\).
\end{eqnarray}
Here $S_G$, $S_C$ and $J$ indicate contributions from soft wide-angle, soft collinear and hard collinear regions, respectively. We here furthermore used the short-hand notation 
\begin{equation}
L^K_{\mu}=\log\(\frac{Q^2}{\mu^2}\)+2p^{(2)}_K\log 2-\left(p^{\left(z_{\mathrm{cut}}\right)}_K-p^{\left(\bar{N}\right)}_K\right)\log x_{L}.
\end{equation}

\noindent
The explicit coefficients needed for NLL accuracy read
\begin{eqnarray}
\Gamma_J &=& 2 C_F \Gamma_{\rm cusp},\\
\Gamma_{S_G} &=& \frac{2}{\beta+1} C_F \Gamma_{\rm cusp},\\
\Gamma_{S_C} &=&-\frac{\beta+2}{\beta+1} C_F \Gamma_{\rm cusp},\\
\gamma^{(0)}_{J}&=&\frac{3}{2}C_F,\\
\gamma^{(0)}_{S_G}=\gamma^{(0)}_{S_C}&=&0,
\end{eqnarray}
where
\begin{equation}
\Gamma_{\rm cusp}=\sum_{n=0}^{\infty}\Gamma^{(n)}_{\rm cusp} \(\frac{\as}{\pi}\)^{n+1}, \quad \text{with} \quad
\Gamma^{(0)}_{\rm cusp}=1,\; \Gamma^{(1)}_{\rm cusp}=\frac{C_A}{2} \(\frac{67}{18}-\frac{\pi^2}{6}\)-\frac{5}{9}T_R n_f.
\end{equation}
Finally, the power coefficients are given by
\begin{eqnarray}
p^{(\bar{N})}_{J}&=-\frac{1}{2},&\quad p^{(\zc)}_{J}=0\\
p^{(\bar{N})}_{S_G}&=0,&\quad p^{(\zc)}_{S_G}=1\\
p^{(\bar{N})}_{S_C}&=\frac{\beta+1}{\beta+2},&\quad p^{(\zc)}_{S_C}=\frac{1}{\beta+2},
\end{eqnarray}
and $p^{(2)}_K=0$ for all $K$.

\subsection{Treatment of the transition point}

A significant difference with respect to the approach of Ref.~\cite{Baron:2018nfz} is the observation that,
if we limit ourselves to NLL, as we do in this study, the transition point, marking the boundary between
the groomed and ungroomed regions, can be taken at $\tau=\zc 2^{\beta/2}$. Thus, we naturally resum logarithms
of $x_L\zc 2^{\beta/2}$ instead of logarithms of $x_L\zc/2$ and consequently all coefficients for
the logarithms of 2 vanish, i.e.\ $p^{(2)}_K=0$ for all $K$. 

We also include an additional correction associated with a
discontinuity of the differential soft-drop thrust distribution at
$\tau=z_\beta\equiv\zc 2^{\beta/2}$ at NLL accuracy, which was not
included in Ref.~\cite{Baron:2018nfz}.
To identify the origin of the discontinuity, let us start with the
expression for the resummed (cumulative) as given e.g.\ in
Ref.~\cite{Banfi:2004yd}:
\begin{equation}\label{eq:caesar-base}
  \Sigma_{\rm res}(\tau)
  =\lim_{\epsilon\to0}\sum_{n=0}^{\infty}\frac{1}{n!}\prod_{i=1}^{n}\int_{\epsilon\tau}^{\tau}\frac{d\tau_{i}}{\tau_{i}}R^{\prime}\left(\tau_{i}\right)\exp\left[-R\left(\epsilon\tau\right)\right]\Theta\left(\tau-\sum_{i=1}^{n}\tau_{i}\right),
\end{equation}
where the exponent is related to the one presented in the previous subsection as $R\(\tau\)=R\(\lambda_{\bar{N}},\lambda_{\zc}\)|_{N\to\tau}$.
One can then use the following expansions:
\begin{align}\label{eq:expansion-normal}
  R(\epsilon\tau) & = R(\tau)
                    + \log\left(\frac{1}{\varepsilon}\right)\,R^\prime(\tau) 
                    + \frac{1}{2} \log^2\left(\frac{1}{\varepsilon}\right)\,R^{\prime\prime}(\tau) 
                    + {\cal{O}}(R^{\prime\prime\prime}(\tau)), \\
  R^\prime(\tau_i) & = R^\prime(\tau)
                     + \log\left(\frac{\tau}{\tau_i}\right)\,R^{\prime\prime}(\tau) 
                     + {\cal{O}}(R^{\prime\prime\prime}(\tau)).
\end{align}
At NLL accuracy, we can only keep the terms proportional to $R(\tau)$
and $R^\prime(\tau)$. Eg.~(\ref{eq:caesar-base}) can then be evaluated
and one gets
\begin{equation}\label{eq:NLL-base}
  \Sigma_{\rm res}^{\text{(NLL)}}(\tau)
  = \frac{\exp\left[-R\left(\tau\right)-\gamma_{E}R^{\prime}\left(\tau\right)\right]}
  {\Gamma\left(1+R^{\prime}\left(\tau\right)\right)}.
\end{equation}
For soft-drop thrust, the issue is that $R^{\prime\prime}(\tau)$ is
discontinuous at $\tau=z_\beta$, meaning that if one computes the
differential distribution by taking the derivative of
\eqref{eq:NLL-base}, one would get a discontinuity at the transition
point $\tau=z_\beta$.

Since this contribution is proportional to $R^{\prime\prime}(\tau)$,
it is formally NNLL in $\log(1/\tau)$. Here, we want to extract from
the NNLL corrections the part responsible for the discontinuity and
include it as a transition-point correction.
To do this, we now include the terms proportional to
$R^{\prime\prime}(\tau)$ in~\eqref{eq:expansion-normal}. We first
consider the case where $\tau<z_\beta$ for
which~(\ref{eq:caesar-base}) becomes
\begin{eqnarray*}
	\left.\Sigma_{\rm res}(\tau)\right|_{\tau<\zcb} & = & \lim_{\epsilon\to0}\sum_{n=0}^{\infty}\frac{1}{n!}\prod_{i=1}^{n}\int_{\epsilon\tau}^{\tau}\frac{d\tau_{i}}{\tau_{i}}\left[R^{\prime}\left(\tau\right)+\log\left(\frac{\tau}{\tau_{i}}\right)R^{\prime\prime}\left(\tau\right)\right]\\
	&  & \times\exp\left[-R\left(\tau\right)-\log\left(\frac{1}{\epsilon}\right)R^{\prime}\left(\tau\right)-\frac{1}{2}\log^{2}\left(\frac{1}{\epsilon}\right)R^{\prime\prime}\left(\tau\right)\right]\Theta\left(\tau-\sum_{i=1}^{n}\tau_{i}\right).
\end{eqnarray*}
At the accuracy of interest, it is sufficient to include a single
correction proportional to $R^{\prime\prime}(\tau)$ and one obtains
after a few trivial manipulations
\begin{equation}\label{eq:trcorrection-below}
\left.\Sigma_{\rm res}(\tau)\right|_{\tau<\zcb} = \frac{\exp\left[-R\left(\tau\right)-\gamma_{E}R^{\prime}\left(\tau\right)\right]}{\Gamma\left(1+R^{\prime}\left(\tau\right)\right)}\left\{ 1+\int_{0}^{1}\frac{dx_{0}}{x_{0}}\log\left(\frac{1}{x_{0}}\right)R^{\prime\prime}\left(\tau\right)\left[\left(1-x_{0}\right)^{R^{\prime}\left(\tau\right)}-1\right]\right\}.
\end{equation}
The integration over $x_0$ in the above expression can be performed
and one recovers the soft-collinear NNLL corrections associated with
multiple-emission (cf.\ Ref.~\cite{Banfi:2014sua}).

When $\tau>z_\beta$, one has to be a bit more careful as the
expansions in~(\ref{eq:expansion-normal}) can involve crossing the
transition point where $R^{\prime\prime}$ is discontinuous. In
particular, for $\tau_i<z_\beta<\tau$ and for
$\epsilon\tau<z_\beta<\tau$, one should instead use
\begin{eqnarray}
\tau_{i}<\zcb<\tau: &  & R^{\prime}\left(\tau_{i}\right)=R^{\prime}\left(\tau\right)+\log\left(\frac{\tau}{\zcb}\right)R^{\prime\prime}\left(\tau\right)+\log\left(\frac{\zcb}{\tau_{i}}\right)R^{\prime\prime}\left(\zcb\right)+{\cal O}\left(R^{\prime\prime\prime}\left(\tau\right)\right),\\
\epsilon\tau<\zcb<\tau: &  & R\left(\epsilon\tau\right)=R\left(\tau\right)+\log\left(\frac{1}{\epsilon}\right)R^{\prime}\left(\tau\right)\\
&  & +\frac{1}{2}\log^{2}\left(\frac{\tau}{\zcb}\right)R^{\prime\prime}\left(\tau\right)+\log\left(\frac{\tau}{\zcb}\right)\log\left(\frac{\zcb}{\epsilon\tau}\right)R^{\prime\prime}\left(\tau\right)+\frac{1}{2}\log^{2}\left(\frac{\zcb}{\epsilon\tau}\right)R^{\prime\prime}\left(\zcb\right)+{\cal O}\left(R^{\prime\prime\prime}\left(\tau\right)\right).\nonumber 
\end{eqnarray}
Following the same procedure as for $\tau<z_\beta$ leads to
\begin{eqnarray}\label{eq:trcorrection-above}
\left.\Sigma_{\rm res}(\tau)\right|_{\tau>\zcb} & = & \frac{\exp\left[-R\left(\tau\right)-\gamma_{E}R^{\prime}\left(\tau\right)\right]}{\Gamma\left(1+R^{\prime}\left(\tau\right)\right)}\left\{ 1+\int_{\zcb/\tau}^{1}\frac{dx_{0}}{x_{0}}\log\left(\frac{1}{x_{0}}\right)R^{\prime\prime}\left(\tau\right)\left[\left(1-x_{0}\right)^{R^{\prime}\left(\tau\right)}-1\right]\right.\\
&  & \left.+\int_{0}^{\zcb/\tau}\frac{dx_{0}}{x_{0}}\left[\log\left(\frac{\tau}{\zcb}\right)\left\{ R^{\prime\prime}\left(\tau\right)-R^{\prime\prime}\left(\zcb\right)\right\} +\log\left(\frac{1}{x_{0}}\right)R^{\prime\prime}\left(\zcb\right)\right]\left[\left(1-x_{0}\right)^{R^{\prime}\left(\tau\right)}-1\right]\right\} .\nonumber
\end{eqnarray}
One can further simplify these expressions by keeping only the
contributions associated with the transition point. This is
equivalent to subtracting the NNLL correction below the transition
point, i.e.\ the correction that a standard NNLL calculation would
give, from both~(\ref{eq:trcorrection-below})
and~(\ref{eq:trcorrection-above}). One then obtains our final expressions
\begin{eqnarray}\label{eq:final-transition-correction}
\left.\Sigma_{\rm res}(\tau)\right|_{\tau<\zcb} & = & \frac{\exp\left[-R\left(\tau\right)-\gamma_{E}R^{\prime}\left(\tau\right)\right]}{\Gamma\left(1+R^{\prime}\left(\tau\right)\right)},\\
\left.\Sigma_{\rm res}(\tau)\right|_{\tau>\zcb} & = & \frac{\exp\left[-R\left(\tau\right)-\gamma_{E}R^{\prime}\left(\tau\right)\right]}{\Gamma\left(1+R^{\prime}\left(\tau\right)\right)}\left\{ 1+\int_{0}^{\zcb/\tau}\frac{dx_{0}}{x_{0}}\left[\log\left(\frac{\tau}{\zcb}\right)\left[ R^{\prime\prime}\left(\tau\right)-R^{\prime\prime}\left(\zcb\right)\right] \right]\left[\left(1-x_{0}\right)^{R^{\prime}\left(\tau\right)}-1\right]\right\} \nonumber \\
& = & \frac{\exp\left[-R\left(\tau\right)-\gamma_{E}R^{\prime}\left(\tau\right)\right]}{\Gamma\left(1+R^{\prime}\left(\tau\right)\right)}\exp\left\{ \int_{0}^{\zcb/\tau}\frac{dx_{0}}{x_{0}}\left[\log\left(\frac{\tau}{\zcb}\right)\left[ R^{\prime\prime}\left(\tau\right)-R^{\prime\prime}\left(\zcb\right)\right] \right]\left[\left(1-x_{0}\right)^{R^{\prime}\left(\tau\right)}-1\right]\right\}, \nonumber
\end{eqnarray}
where the exponentiation of this contribution holds at NLL accuracy
for both cumulative cross section and the differential distribution.
Here the universal contribution is the usual Mellin inversion
contribution, whereas an additional is included above the transition
point.

One can show that the above result is, as expected, continuous at
$\tau=z_\beta$. It is however interesting to comment on how this
happens. The correction in the curly bracket
in~\eqref{eq:final-transition-correction} are contributing only at the
NNLL accuracy, and one can view the $\log(z_\beta/\tau)$ in the
exponent as being small. However, when taking the derivative with respect to
$\log(1/\tau)$, it gives a correction which exactly compensates the
divergence in~(\ref{eq:NLL-base}). This extra dependence of the
coefficient of $R^{\prime\prime}(\tau)$ contrasts with the standard
case (cf.~Eq.~(\ref{eq:trcorrection-below})) where it does not depend on $\tau$ at NNLL accuracy.

\subsection{End-point correction}\label{sec:endpoint}
Here we briefly describe our treatment to ensure that the resummed cross section and its expansion respect the
kinematic end-point $\tau_{\rm max}$ of the fixed-order calculation, we thereby follow the procedure given in
Refs.~\cite{Catani:1992ua,Jones:2003yv}. The primary modification is an alteration of the argument of the logarithm
\begin{equation}
\log\left(x_{L}\tau\right)\to-\frac{1}{p}\log\left(\frac{1}{\left(x_{L}\tau\right)^p}-\frac{1}{\left(x_{L}\tau_{\mathrm{max}}\right)^p}+1\right)=\log\bar{\tau},
\end{equation}
where the parameter $p$ determines the slope of the approach to the end-point. 

However, the modification of the logarithm is not enough in order to ensure for the derivative of the expansion to
approach $0$. To implement this an additional contribution is included in the resummed exponential
\begin{equation}
\Sigma_{\rm res}\to\Sigma_{\rm res} \exp\left[-\left(\frac{\tau}{\tau_{\mathrm{max}}}\right)^p \tilde{R}^{\prime}\left(\tau\to\tau_{\mathrm{max}}\right)\log\bar{\tau}\right],
\end{equation}
where $\tilde{R}$ includes the transition-point and multiple-emission effects.
The derivative here is taken with respect to $\log\bar{\tau}$.

The value of the end-point depends on the accuracy of the fixed-order calculation which we match to. At NLO precision this
is given by $\tau_{\mathrm{max}}=0.4225$~\cite{Jones:2003yv}.

\section{An analytic model for hadronisation corrections}\label{sec:hadr}

In this section the analytic hadronisation model is described in detail. The derivation follows the approach of \cite{Dasgupta:2007wa},
which was applied to mMDT and soft drop in \cite{Dasgupta:2013ihk,Marzani:2017kqd}.
\subsection{Non-perturbative effects on plain thrust}
For plain thrust non-perturbative corrections result in a simple shift of the observable value. 
To derive this shift we evaluate the observable for a $2\to2$ process with one
additional non-perturbative emission $k$ off the hard legs $1$ and $2$. 
We find it convenient to align the $z$ axis with one of the final-state
particles, so the kinematics for this computation are given by 
\begin{eqnarray}
p_2&=&E_2\(1,0,0,1\)\,,\nonumber\\
k&=&E_k\(1,0,\sqrt{1-c_{2k}^2},c_{2k}\)\,,
\end{eqnarray}
where $c_{2k}=\cos\theta_{2k}$, with $\theta_{23}$ the angle between
leg 2 and the soft non-perturbative emission $k$, while $p_1=q-p_2-k$,
with $q=Q(1,0,0,0)$.  Exploiting energy momentum conservation and
on-shell relations, we can express the energies $E_1$ and $E_2$ as a
function of the energy of the non-perturbative emission $E_k$ and
$c_{2k}$
\begin{align}
E_1&= \frac{Q^2-2(1-c_{2k})E_k(Q-E_k)}{2\(Q-E_k\(1-c_{2k}\)\)}, \\
E_2&=\frac{Q\(Q-2E_k\)}{2\(Q-E_k\(1-c_{2k}\)\)}\,.
\end{align}
This results in a contribution to the thrust 
\begin{align}
\tau\(k\)=\delta\tau^{+} \(k\) = 1-\frac{2E_1}{Q} &=   \frac{E_{k}\left(Q-2E_{k}\right)\left(1-c_{2k}\right)}{Q\left(Q-E_{k}\left(1-c_{2k}\right)\right)}\,,&&\text{when}\;\; c_{2k}>-\frac{E_{k}}{Q-E_{k}}\,,\\
\tau\(k\)=\delta\tau^{-} \(k\) = 1-\frac{2E_2}{Q} &= \frac{E_{k}\left(1+c_{2k}\right)}{\left(Q-E_{k}\left(1-c_{2k}\right)\right)}\,, &&\text{when}\;\; c_{2k}<-\frac{E_{k}}{Q-E_{k}}\,,
\end{align}
where the two cases correspond to the hemisphere in which the non-perturbative emission resides.
Making use of the eikonal rules we can write down an integral for the expectation value of the shift
as a result of this non-perturbative emission:
\begin{equation}
\left\langle \delta\tau\right\rangle_h =C_F\int dE_{k}E_{k}dc_{2k}\frac{\as\left(k_{t,12}\right)}{2\pi}\frac{p_{1}\cdot p_{2}}{\left(p_{1}\cdot k\right)\left(p_{2}\cdot k\right)}\delta\tau\left(k\right)\,,
\end{equation}
where the scale for $\as$~\cite{Catani:1990rr} is given by
\begin{equation}
k_{t,12}^{2}=2\frac{\left(p_{1}\cdot k\right)\left(p_{2}\cdot k\right)}{p_{1}\cdot p_{2}}
\end{equation}
and 
\begin{align}\label{eq:del_tau}
\delta\tau\left(k\right)&=\delta\tau^{+}\left(k\right)\Theta\left(c_{2k}+\frac{E_{k}}{Q-E_{k}}\right)+\delta\tau^{-}\left(k\right)\left[1-\Theta\left(c_{2k}+\frac{E_{k}}{Q-E_{k}}\right)\right]
\nonumber\\
&=
\delta\tau^{-}\left(k\right)+\Theta\left(c_{2k}+\frac{E_{k}}{Q-E_{k}}\right) \left[\delta\tau^{+}(k)-\delta\tau^{-}(k)\right]
\,.
\end{align}
It can be shown~\cite{Dasgupta:2007wa} that the first contribution is
power-suppressed with the exception of the collinear divergence, which will anyway cancel. Therefore we shall focus on the difference
\begin{equation}
\delta\tau^{+}(k)-\delta\tau^{-}(k)=-\frac{2E_{k}}{Q}\frac{Qc_{2k}+E_{k}\left(1-c_{2k}\right)}{Q-E_{k}\left(1-c_{2k}\right)}\,.
\end{equation}
In addition, with $k_{t,12}$ being the argument of $\as$, we change the integration variable to
\begin{equation}
k_{t,12}=E_{k}\sqrt{1-c_{2k}^{2}}\frac{Q}{Q-E_{k}\left(1-c_{2k}\right)}\,,
\end{equation}
which, for simplicity, we denote as $k_t$ in what follows. This results in the final integral
\begin{equation}
\left\langle \delta\tau\right\rangle _{h} = -C_F\int_{0}^{\mu_{I}}\frac{dk_{t}}{Q}\frac{\delta\as\left(k_{t}\right)}{\pi}\int_{-1}^{1}dc_{2k}\left[2c_{2k}\left(1-c_{2k}^{2}\right)^{-3/2}\Theta\left(c_{2k}\right)+{\cal O}\left(k_{t}/Q\right)\right]\,,
\end{equation}
where we have performed an expansion with $k_{t}$ as the softness parameter.
$\delta\as\left(k_{t}\right)$ denotes the difference between the non-perturbative and the
perturbative description of the strong coupling. The angular integral can easily be performed,
with the collinear divergence cancelling against the first term
in \eqref{eq:del_tau}, and we can redefine the $k_{t}$ integral
\begin{equation}
\left\langle \delta\tau\right\rangle _{h}=2C_{F}\frac{\Lambda\left(\mu_{I}\right)}{Q}\,,
\end{equation}
where we have introduced
\begin{equation}
\Lambda\left(\mu_{I}\right)=\int_{0}^{\mu_{I}}\frac{dk_{t}}{\pi}\delta\as\left(k_{t}\right)\,.
\end{equation}
The actual fitted parameter is defined as
\begin{equation}
\Omega=C_F \Lambda\left(\mu_{I}\right)\,.
\end{equation}

\subsection{Non-perturbative effects on soft-drop thrust}\label{sec:hadr_sd}

Now that we have reviewed the effect of a non-perturbative emission on the value of thrust we can study the groomed distribution. In general a non-perturbative emission for the $2\to2$ case will be groomed away leaving no change in thrust unless the emission is hard enough and very collinear (resulting in a small $k_{t}$ but large enough energy).

This can be computed by including the soft-drop restriction for the non-perturbative emission:
\begin{equation}
k_t>\zc\frac{Q}{2}\(1-c_{2k}\)^{(1+\beta)/2}\(1+c_{2k}\)^{1/2},
\end{equation}
where the small $k_t$ approximation has been used. This can be rewritten as a condition on $c_{2k}$ under the assumption that $1+c_{2k}$ is close to $2$, i.e.
\begin{equation}
c_{2k}>1-\(\frac{\sqrt{2}k_t}{\zc Q}\)^{2/(\beta+1)}.
\end{equation}
The integral will now be over $\delta\tau^+\(k\)$ instead of the difference as there is no shift if the emission is groomed. Performing the angular integral results in a contribution:
\begin{equation}
\left\langle \delta\tau\right\rangle _{h,SD}=2^{(\beta+2)/(2(\beta+1))}C_{F}\zc^{-1/(\beta+1)}\frac{\Lambda_{SD}\left(\mu_{I},\beta\right)}{Q^{(\beta+2)/(\beta+1)}},
\end{equation}
including a factor 2 for both hemispheres where
\begin{equation}
\Lambda_{SD}\left(\mu_{I},\beta\right)=\int_{0}^{\mu_{I}}\frac{k_t^{1/(\beta+1)}dk_{t}}{\pi}\delta\as\left(k_{t}\right).
\end{equation}
This effect is suppressed in $k_{t}$. Note that the limit $\beta\to\infty$ does not work in this case as there are contributions of the type $2-\(\sqrt{2}k_t/\(\zc Q\)\)^{2/(\beta+1)}$, which approach 1 for $\beta\to\infty$ but result in a factor 2 in the small $k_t$ limit.

A contribution which is not suppressed originates from a $2\to n$ process where the non-perturbative emission is protected by
larger-angle hard emissions, which pass soft drop. This is given by:
\begin{equation}
\left\langle \delta\tau\right\rangle_h =C_{F}\int dE_{k}E_{k}dc_{23,k}\frac{d\phi_{k}}{2\pi}\frac{\as\left(k_{t,ij}\right)}{2\pi}\frac{p_{1}\cdot p_{23}}{\left(p_{1}\cdot k\right)\left(p_{23}\cdot k\right)}\delta\tau^{+}\left(k\right)\Theta\left(c_{23,k}-c_{23}\right),
\end{equation}
where we apply the same shift as for the $2\to2$ kinematics with an
angular constraint now set by the angle $\theta_{23}$ between the
emissions passing the soft-drop condition. Here, the non-perturbative emission is constrained by the angle between the two hard particles in one hemisphere, $c_{23,k}>c_{23}$ . This condition can be translated to a restriction based on a value of thrust. Here we make use of the same kinematics as in last subsection with $k\to p_3$, thus
\begin{equation}
\tau=\frac{E_{2}\left(Q-2E_{2}\right)\left(1-c_{23}\right)}{Q\left(Q-E_{2}\left(1-c_{23}\right)\right)},
\end{equation}
resulting in
\begin{equation} \label{eq:c23oftau}
c_{23}=1-\frac{2\tau}{\left(1-z\right)\left(z+\tau\right)},
\end{equation}
with $z=1-2E_2/Q$.

We have once again
\begin{equation}
k_{t,1,23}=E_{k}\sqrt{1-c_{23,k}^{2}}\frac{Q}{Q-E_{k}\left(1-c_{23,k}\right)}.
\end{equation}
This results in the integral
\begin{eqnarray}
\left\langle \delta\tau\right\rangle _{h} & = & C_{F}\int\frac{dk_{t}}{Q}\frac{\delta\as\left(k_{t}\right)}{2\pi}dx\frac{d\phi_{k}}{2\pi}\frac{\tau^{1/2}\left(1-z\right)\left(\tau+z\right)}{x^{1/2}\left(\tau\left(1-x-z\right)+\left(1-z\right)z\right)^{3/2}}+{\cal O}\left(k_{t}/Q\right)\nonumber \\
& = & C_{F}\frac{\Lambda\left(\mu_{I}\right)}{Q}\frac{\tau^{1/2}}{\left(z\left(1-z-\tau\right)\right)^{1/2}}=C_{F}\frac{\Lambda\left(\mu_{I}\right)}{Q}\frac{\tau^{1/2}}{\left(z\left(1-z\right)\right)^{1/2}}+{\cal O}\left(\tau\right),
\end{eqnarray}
with $1-c_{23,k}=x\left(1-c_{23}\right)$. 

The mean value of this shift can be obtained by averaging the above result with the appropriate QCD splitting function. Here the integration boundaries will have a transition point in thrust:
\begin{equation}
\left\langle \left(z\left(1-z\right)\right)^{-1/2}\right\rangle =\frac{\int_{Z_{SD}-\tau}^{1-Z_{SD}}dz\left(z\left(1-z\right)\right)^{-1/2}P_{qg}\left(z\right)}{\int_{Z_{SD}-\tau}^{1-Z_{SD}}dzP_{qg}\left(z\right)},
\end{equation}
for $\tau<\zc/2$, and
\begin{equation}
\left\langle \left(z\left(1-z\right)\right)^{-1/2}\right\rangle =\frac{\int_{\tau}^{1-2\tau}dz\left(z\left(1-z\right)\right)^{-1/2}P_{qg}\left(z\right)}{\int_{\tau}^{1-2\tau}dzP_{qg}\left(z\right)},
\end{equation}
for $\tau>\zc/2$ with $Z_{SD}=\zc^{2/\left(2+\beta\right)}\left(2\tau\right)^{\beta/\left(2+\beta\right)}$.
Finally for $\tau>1/3$ both these integrals approach 0 simultaneously and we make use of the limit
\begin{equation}
\left\langle \left(z\left(1-z\right)\right)^{-1/2}\right\rangle = \frac{3}{\sqrt{2}},
\end{equation}
however this is not relevant to the fitting range.

\subsection{Shift in energy}

In addition to shifting the value of thrust it is also possible to shift the energy of an emitted gluon downwards to below the needed energy fraction in order to pass the grooming condition. Effectively this shift in the energy can be interpreted as a shift in $\zc$
\begin{equation}
\delta \zc=-2\left(\frac{\(z+\tau\)\(1-z\)}{2\tau}\right)^{\beta/2}\frac{\delta E}{Q},
\end{equation}
where we have made use of $E_2=\(1-z\)Q/2$ and $E_3=\(z+\tau\)Q/2$ and
we have assumed the angle is fixed and determined by the value of
thrust. Here we can compute the energy difference using the same
techniques than those applied to derive the shift in thrust. Once again we will make use of the observable difference for the $2\to2$ case:
\begin{eqnarray}
\delta E^{+}\(k\) & = & E_{3}+E_{k}-Q/2=\frac{E_{k}\left(Q-2E_{k}\right)\left(1-c_{3k}\right)}{2\left(Q-E_{k}\left(1-c_{3k}\right)\right)},\\
\delta E^{-}\(k\) & = & E_{3}-Q/2=\frac{-E_{k}Q\left(1+c_{3k}\right)}{2\left(Q-E_{k}\left(1-c_{3k}\right)\right)}.
\end{eqnarray}
The first shift corresponds to the case where the particles $k$ and $3$ are clustered together first, whereas the second case assumes $k$ does not cluster with $3$. This results in the integral

\begin{equation}
\left\langle \delta E\right\rangle_{h} =C_{A}\int dE_{k}E_{k}dc_{2k}\frac{\as\left(k_{t,12}\right)}{2\pi}\frac{p_{1}\cdot p_{3}}{\left(p_{1}\cdot k\right)\left(p_{3}\cdot k\right)}\left(\delta E^{+}-\delta E^{-}\right)\Theta\left(c_{3k}-c_{23}\right).
\end{equation}
Note the different colour factor with respect to the previous effect: we now expect the dominant contribution to come from a non-perturbative gluon that shifts the energy of the perturbative one. 
Again we use
\begin{equation}
k_{t,13}=E_{k}\sqrt{1-c_{3k}^{2}}\frac{Q}{Q-E_{k}\left(1-c_{3k}\right)}.
\end{equation}
Resulting in the integral
\begin{eqnarray}
\left\langle \delta E\right\rangle _{h} & = & C_{A}\int dk_{t}\frac{\delta\as\left(k_{t}\right)}{2\pi}dc_{3k}\frac{d\phi_{k}}{2\pi}\frac{2}{\left(1-c_{3k}^{2}\right)^{3/2}}+{\cal O}\left(k_{t}/Q\right)\nonumber \\
& =- & C_{A}\Lambda\left(\mu_{I}\right)\frac{c_{23}}{\sqrt{1-c_{23}^{2}}}=-C_{A}\Lambda\left(\mu_{I}\right)\frac{z\left(1-z-\tau\right)-\tau}{2\sqrt{\tau z\left(1-z-\tau\right)}}.
\end{eqnarray}
Since we know $z\sim Z_{SD}-\tau$ we can fill this into the expression
resulting in
\begin{equation}
\left\langle \delta \zc\right\rangle _{h} = 2C_{A}\left(\frac{Z_{SD}\(1-Z_{SD}+\tau\)}{2\tau}\right)^{\beta/2}\frac{\(Z_{SD}-\tau\)\left(1-Z_{SD}\right)-\tau}{2\sqrt{\tau \(Z_{SD}-\tau\)\left(1-Z_{SD}\right)}}\frac{\Lambda\left(\mu_{I}\right)}{Q}.
\end{equation}
However this only holds for $\tau<\zc/2$, therefore we freeze the
shift for $\tau>\zc/2$ leading to
\begin{equation}
\left\langle \delta \zc\right\rangle _{h} = -2C_{A}\left(1-\zc/2\right)^{\beta/2}\frac{\zc}{2\sqrt{1-\zc}}\frac{\Lambda\left(\mu_{I}\right)}{Q}\,,\quad\text{when}\;\; \tau>\frac{\zc}{2}\,.
\end{equation}

\section{Validation of Hadronisation corrections from Monte Carlo generators}\label{sec:MC-val}

\begin{figure}[!htb]
  \centering
  \includegraphics[width=\textwidth]{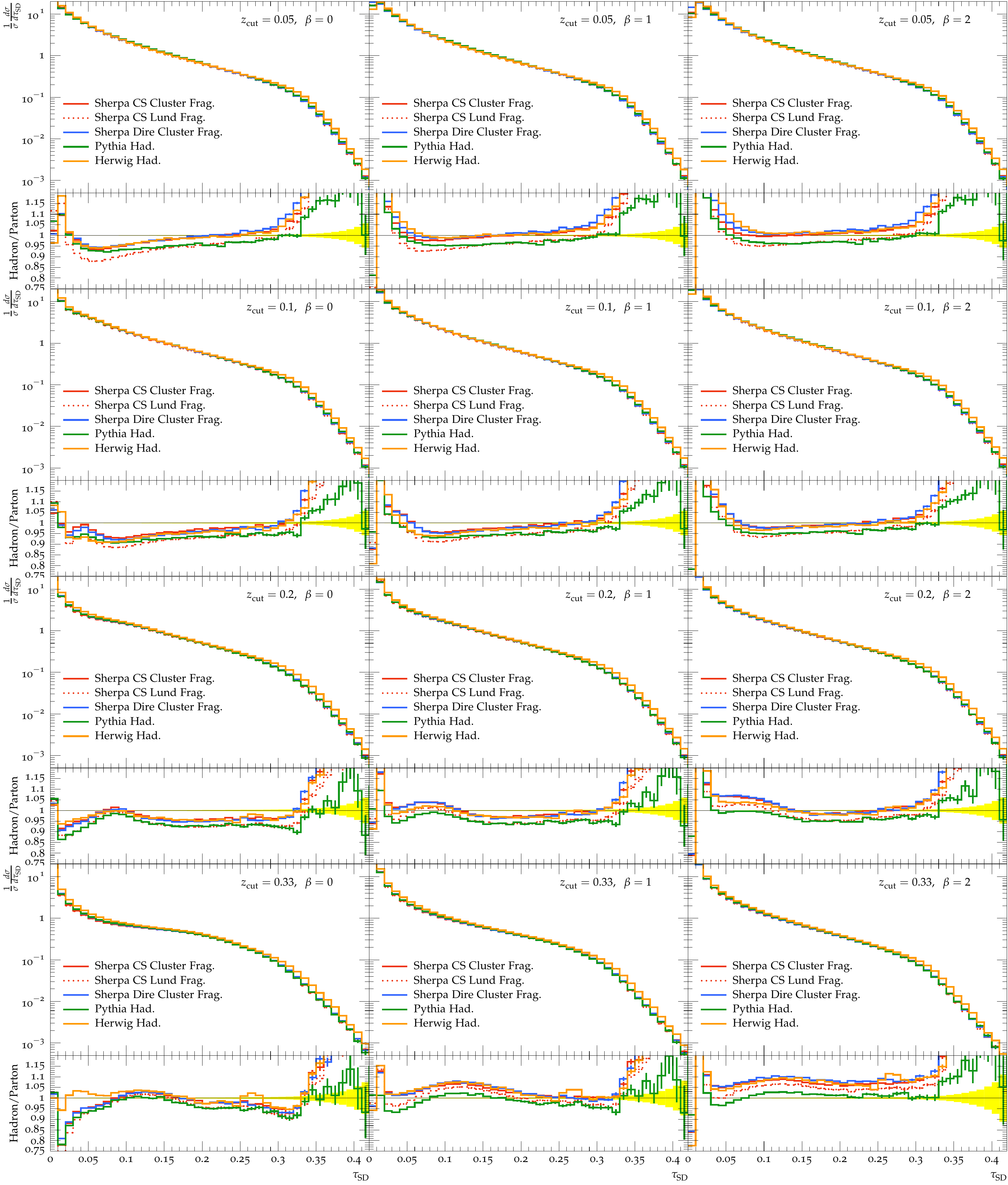}
  \caption{Predictions from general-purpose Monte Carlo generators for soft drop
    thrust with various $\zc$ and $\beta$ values at parton shower
    accuracy. Shown are the nominal distributions at hadron level and the ratios
    of hadron level to the respective underlying parton level
    predictions.}\label{fig:MC-comp} 
\end{figure}

In Fig.~\ref{fig:MC-comp} we compare the predictions of various general
purpose Monte Carlo generators with different hadronisation models for the
various choices of $z_\mathrm{cut}$ and $\beta$. Being mainly interested
in modifications due to different fragmentation models, we perform this
comparison for LO $e^+e^- \to 2j$ matrix elements with parton showers
attached as the perturbative input for all generators.
In addition to the \sherpa\ setups described in the main text, we use
\sherpa~with the \dire\ parton cascade \cite{Hoche:2015sya} thus changing the
underlying showering algorithm but still using cluster hadronisation model.
We find relatively mild changes in the predicted hadronisation
corrections with respect to using the dipole shower.

We furthermore compare to \herwig\ version 7.1.4 \cite{Bahr:2008pv, Bellm:2015jjp}
and \pythia\ version 8.235 \cite{Sjostrand:2006za, Sjostrand:2007gs}. 
For \herwig\ we use the angular-ordered parton shower in conjunction with its
cluster fragmentation model~\cite{Webber:1983if}.
\pythia\ implements a transverse momentum ordered parton shower supplemented with
the Lund fragmentation model. Both hadronisation models are used with their
respective default tuning parameters \cite{Gieseke:2012ft, Reichelt:2017hts,
  Corke:2010yf}.
The effect of soft-drop grooming on the thrust distribution is modelled
consistently between the generators. We observe that the hadronisation
corrections, taken from ratios between the nominal predictions of the generators
to their respective parton level, are at most about 10\% in the relevant range
for most soft-drop parameters, irrespective of the generator used. For all but
the most extreme choice of $\zc=0.33$, the range between the default
\sherpa~dipole shower with cluster or string hadronisation cover the
spread of the other generator choices, at least in the default fitting range
of the observable. This justifies the exclusive use of the \sherpa~dipole shower to
generate pseudo data for the $\as$ fits and to determine hadronisation corrections,
with the average of cluster and string model as our default choice and the difference
to cluster fragmentation and the string model as an estimate of the related uncertainty. 

\begin{figure}[!htb]
  \centering
  \includegraphics[width=.49\textwidth]{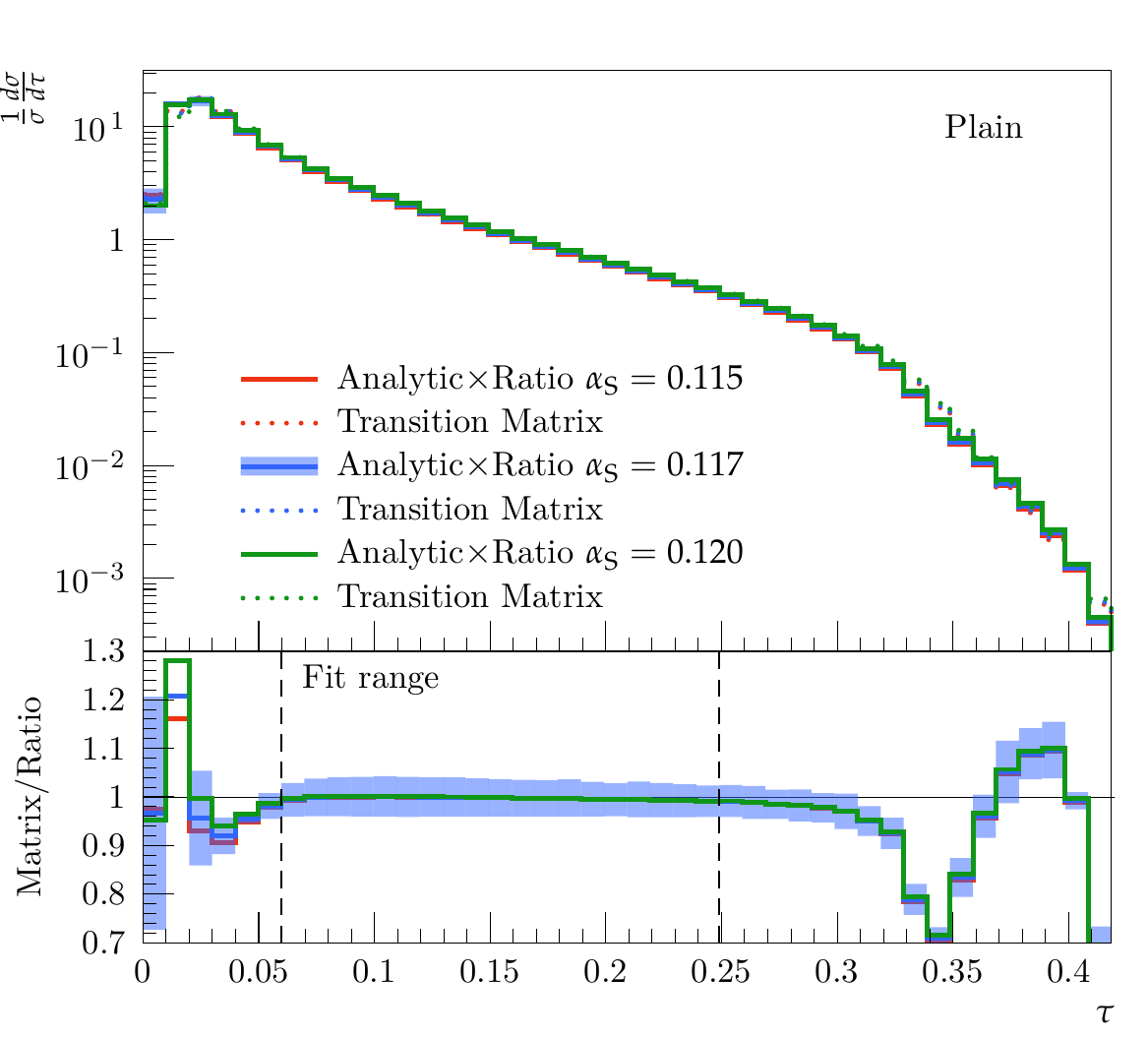}
  \includegraphics[width=.49\textwidth]{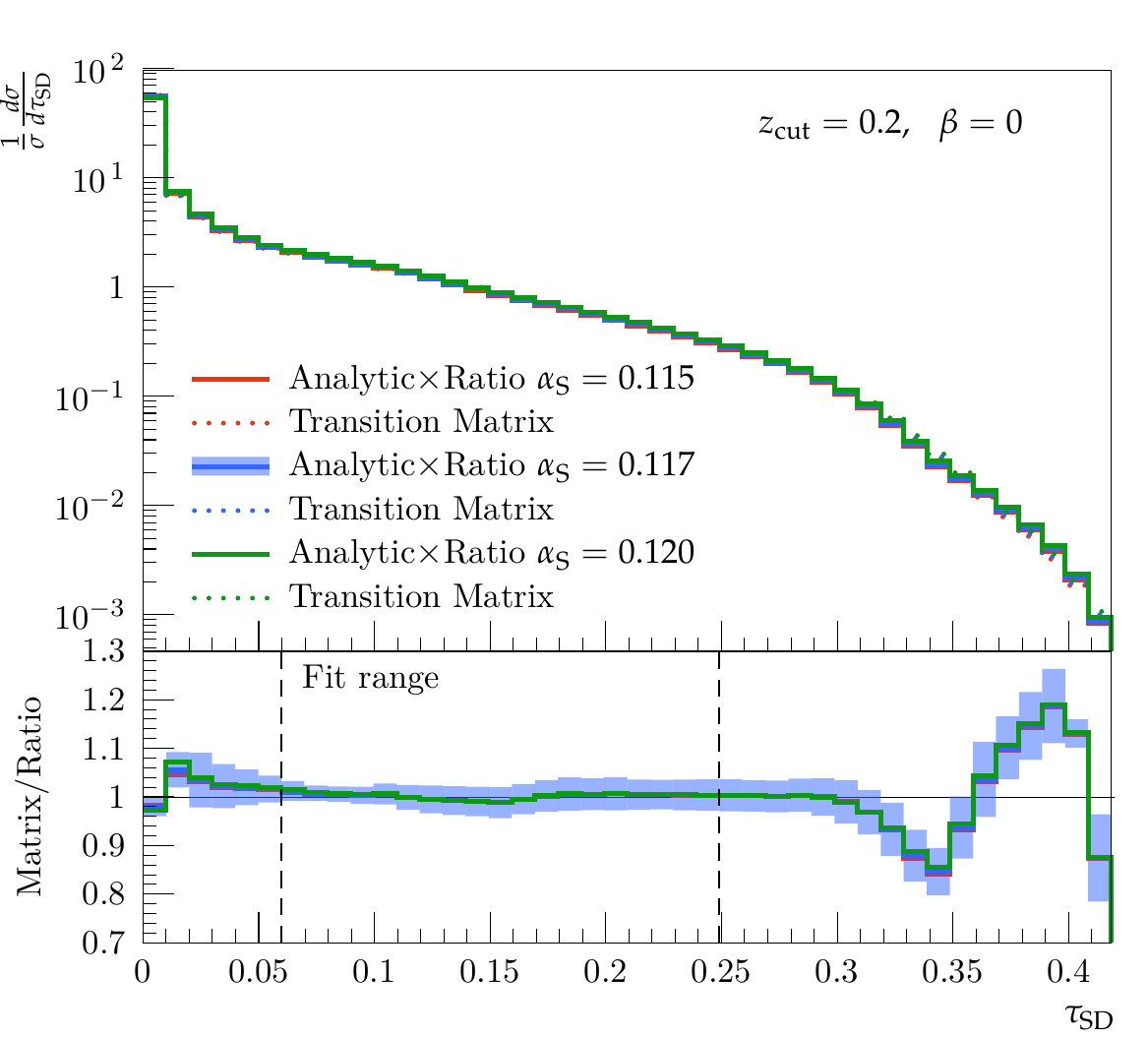}
  \caption{Comparison of different methods to apply hadronisation corrections
    obtained from Monte Carlo to the analytic calculations, for plain thrust (left) and
    soft-drop thrust (right) with $z_\mathrm{cut}=0.2,\beta=0$. The corrections
    are obtained from simulations at parton shower accuracy and are applied
    either by multiplying the analytic calculation at various $\as$ values by
    the ratio from Monte Carlo, as described in the main text (solid) or
    alternatively by applying a bin-by-bin transition matrix obtained from the same Monte
    Carlo run (dotted). The cluster hadronisation model is used in all cases, the
    blue band indicates the uncertainty assigned to the calculation for
    $\as=0.117$ based on the difference to using the Lund model. The bottom
    panel shows the ratio between the distributions using the two methods for
    equal values of $\as$.}\label{fig:HadMethod}
\end{figure}

An alternative way to account for hadronisation corrections other than the ratio between parton and hadron level is to
obtain a bin-by-bin transition matrix from the Monte Carlo and apply 
it to the analytic calculation. We compare the two methods in
Fig. \ref{fig:HadMethod}, for plain thrust and for soft-drop thrust with
parameters $z_\mathrm{cut}=0.2$ and $\beta=0$ (other parameter
choices show a similar behaviour). Although the methods are clearly not
equivalent in the peak region and in the far tail, we find only small numerical
differences in the region where the fit is performed, with
almost no dependence on $\as$. Similarly, small effects are observed for other
parameter choices. We conclude that the difference between these two methods
does not significantly add to the uncertainty assigned to the hadronisation
corrections, and is irrelevant at the overall level of accuracy considered.

\section{Alternative treatment of the resummation uncertainty}\label{sec:unc}
\begin{figure}[h!]
	\centering
	\begin{subfigure}[t]{.48\textwidth}
		\centering
		\includegraphics[width=\textwidth]{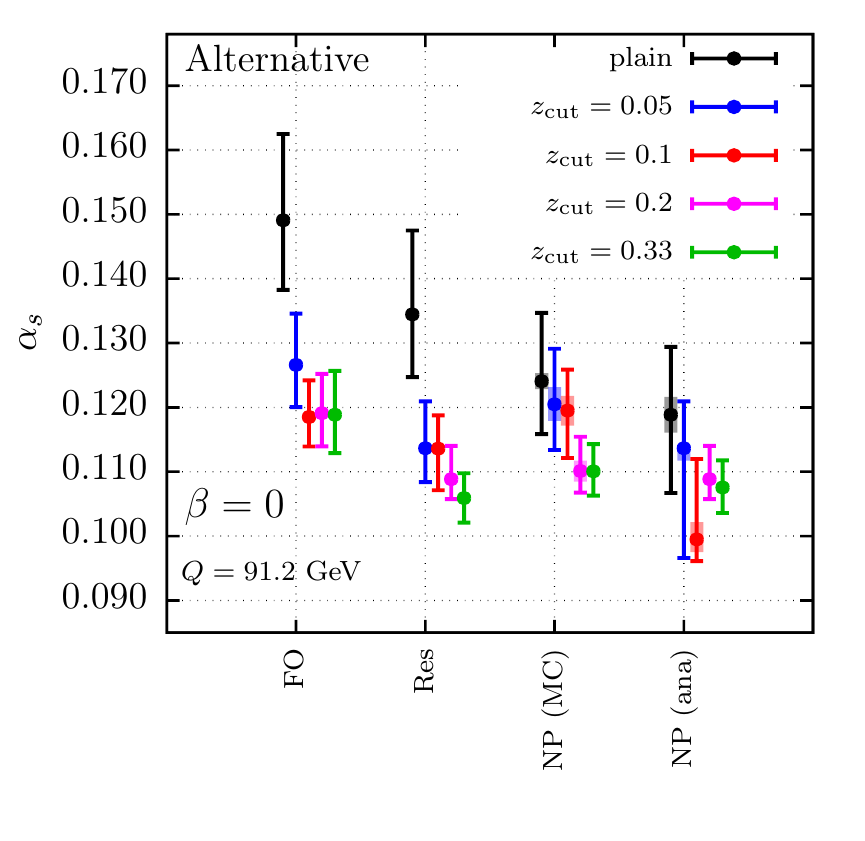}
	\end{subfigure}
	\hfill
	\begin{subfigure}[t]{.48\textwidth}
		\centering
		\includegraphics[width=\textwidth]{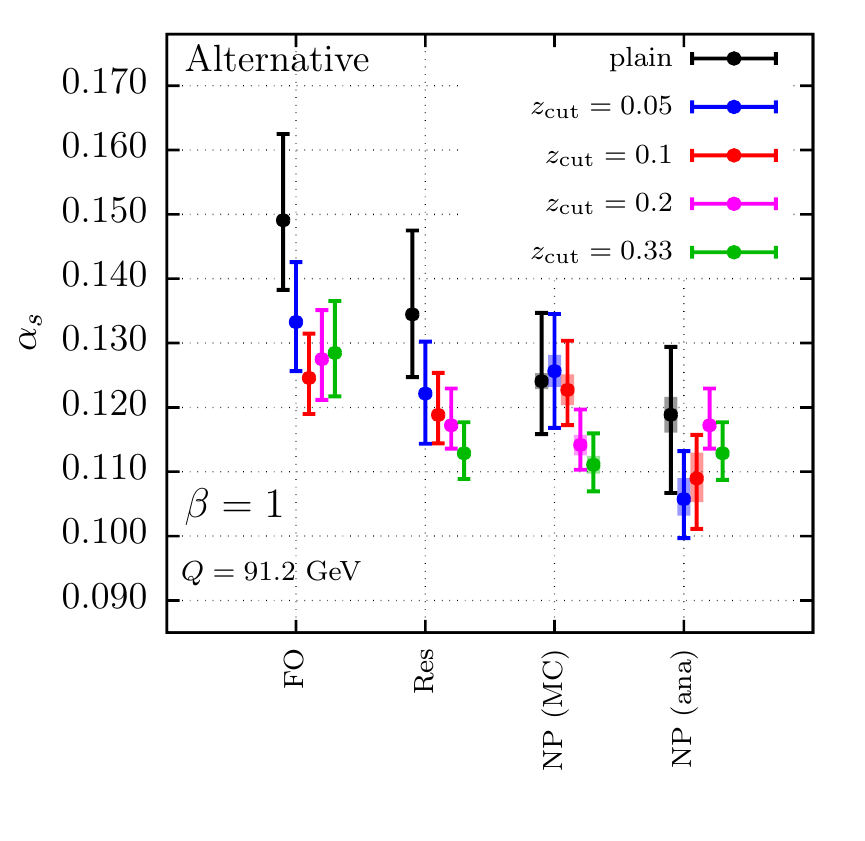}
	\end{subfigure}\\
	\begin{subfigure}[t]{.48\textwidth}
		\centering
		\includegraphics[width=\textwidth]{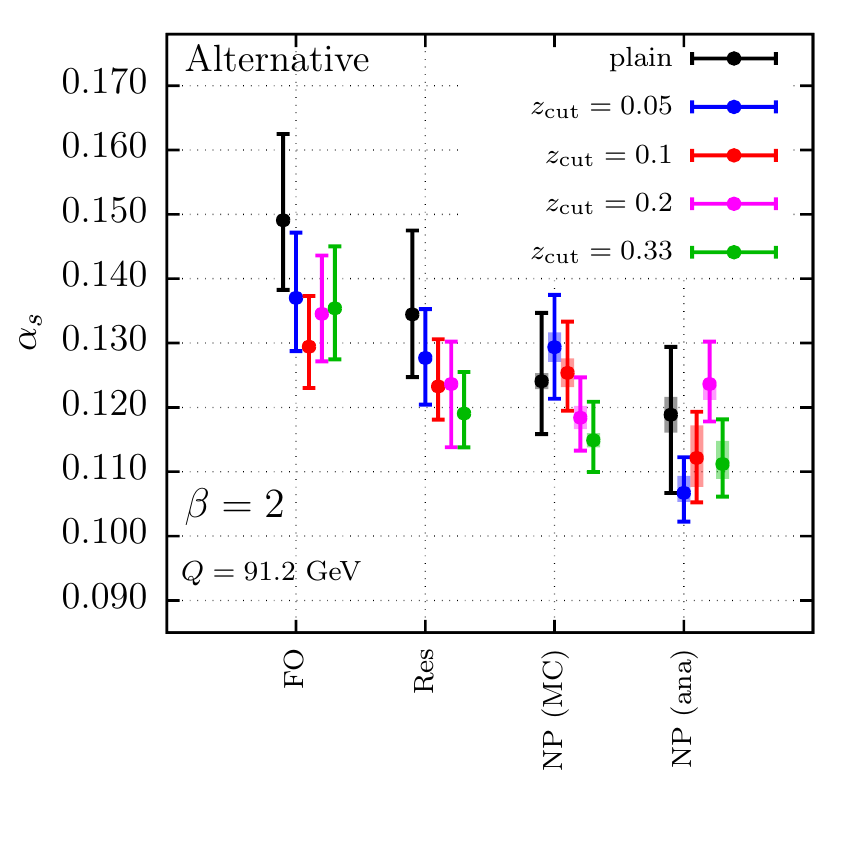}
	\end{subfigure}
	\caption{Similar results to Fig.~\ref{fig:as-fit} making use of an alternative resummation code with a different treatment for the resummation uncertainty.}\label{fig:as-fit_alt}
\end{figure}

In the study we have presented here we have chosen to estimate the theoretical uncertainty
due to missing higher-logarithmic contributions in our resummation by rescaling the arguments
of both the logarithms of thrust and of $\zc$ by an arbitrary factor $x_L$, which we are free
to vary. As an alternative, we can consider to include the rescaling factor in logarithms
of thrust only. In this case the location of the transition point depends on the value of $x_L$. 
In order to test the impact of this another code was developed for the resummation which is
formally equivalent at NLL accuracy, but uses the alternative treatment for estimating 
the resummation uncertainty. In Fig.~\ref{fig:as-fit_alt} the results of the fit using the
alternative approach can be seen. Here the central values are slightly different due to some
formally NNLL differences. However, the main difference is the significantly reduced
uncertainty. In fact the uncertainty shown here is smaller than for plain thrust. We have
decided to adopt a rather conservative approach, i.e.\ using the method described in the main
text and in App.~\ref{sec:mult-trans} as the default choice to estimate the resummation
contribution to the total theoretical uncertainty.

%%%%%%%%%%%%%%%%%%%%%
\phantomsection
\addcontentsline{toc}{section}{References}
\bibliographystyle{jhep}
\bibliography{references}
\end{document}